\documentclass{vldb}

\usepackage{latexsym}
\usepackage{amsfonts}
\usepackage{amsmath}
\usepackage{amssymb}
\usepackage{color}
\usepackage{epsfig}
\usepackage{xspace}
\usepackage{graphicx,epstopdf}
\usepackage{subfigure}
\usepackage{cite}
\usepackage{balance}
\usepackage{textcomp}
\usepackage[rightcaption]{sidecap}
\usepackage{caption}
\usepackage{soul}

\usepackage{blindtext}
\usepackage{etoolbox}

\makeatletter
\patchcmd{\maketitle}{\@copyrightspace}{}{}{}
\makeatother

\newcommand{\myhrule}{\rule[.5pt]{\hsize}{.5pt}}

\newcommand{\eat}[1]{}
\newcommand{\stab}{\rule{0pt}{8pt}\\[-1.6ex]}

\newcommand{\sstab}{\rule{0pt}{8pt}\\[-2.4ex]}

\newcommand{\bi}{\begin{itemize}}
\newcommand{\ei}{\end{itemize}}

\newcommand{\mat}[2]{{\begin{tabbing}\hspace{#1}\=\+\kill #2\end{tabbing}}}

\newcommand{\be}{\begin{enumerate}}
\newcommand{\ee}{\end{enumerate}}
\newcommand{\beqn}{\begin{eqnarray*}}
\newcommand{\eeqn}{\end{eqnarray*}}

\newcommand{\stitle}[1]{\vspace{0.75ex}\noindent{\bf #1}}
\newcommand{\etitle}[1]{\vspace{0.5ex}\noindent{\em \underline{#1}}}

\newcommand{\ie}{\emph{i.e.,}\xspace}
\newcommand{\eg}{\emph{e.g.,}\xspace}
\newcommand{\wrt}{\emph{w.r.t.}\xspace}
\newcommand{\aka}{\emph{a.k.a.}\xspace}
\newcommand{\kwlog}{\emph{w.l.o.g.}\xspace}

\newcommand{\If}{\mbox{\bf if}\ }

\newcommand{\Then}{\mbox{\bf then}\ }

\newcommand{\While}{\mbox{\bf while}\ }

\newcommand{\Do}{\mbox{\bf do}\ }

\newcommand{\For}{\mbox{\bf for}\ }
\newcommand{\Each}{\mbox{\bf each}\ }

\renewcommand{\And}{\mbox{\bf and}\ }

\newcommand{\Continue}{\mbox{\bf continue}\xspace}
\newcommand{\Return}{\mbox{\bf return}\ }

\newcommand{\kw}[1]{{\ensuremath {\mathsf{#1}}}\xspace}

\newcounter{ccc}
\newcommand{\bcc}{\setcounter{ccc}{1}\theccc.}
\newcommand{\icc}{\addtocounter{ccc}{1}\theccc.}

\newcommand{\NP}{{\sc np}\xspace}

\newcommand{\eop}{\hspace*{\fill}\mbox{$\Box$}}     
\newcounter{example}
\renewcommand{\theexample}{\arabic{example}}
\newenvironment{example}{
        \vspace{1ex}
        \refstepcounter{example}
        {\noindent\bf Example \theexample:}}{
        \eop\vspace{1ex}}

\renewcommand{\ni}{\noindent}

\newcommand{\nthesection}{\arabic{section}}
\newcounter{theorem}
\renewcommand{\thetheorem}{\arabic{theorem}}
\newcounter{prop}
\renewcommand{\theprop}{\arabic{theorem}}
\newcounter{property}
\renewcommand{\theprop}{\arabic{theorem}}
\newcounter{lemma}
\renewcommand{\thelemma}{\arabic{theorem}}
\newcounter{cor}
\renewcommand{\thecor}{\arabic{theorem}}
\newenvironment{theorem}{\begin{em}
        \refstepcounter{theorem}
        {\vspace{1.5ex} \noindent\bf  Theorem  \thetheorem:}}{
        \end{em}\eop\vspace{1.0ex}} 
\newenvironment{prop}{\begin{em}
        \refstepcounter{theorem}
        {\vspace{1.5ex}\noindent \bf Proposition \theprop:}}{
        \end{em}\eop\vspace{1.0ex}}

\newenvironment{lemma}{\begin{em}
        \refstepcounter{theorem}
        {\vspace{1.5ex}\noindent\bf Lemma \thelemma:}}{
        \end{em}\eop\vspace{1.0ex}} 
\newcounter{definition}[section]
\renewcommand{\thedefinition}{\nthesection.\arabic{definition}}

\newcounter{alg}[section]
\renewcommand{\thealg}{\nthesection.\arabic{alg}}

\newcounter{arule}
\renewcommand{\thearule}{\arabic{arule}}

\newcounter{claim}
\renewcommand{\theclaim}{\arabic{claim}}

\newenvironment{proofS}{
        \vspace{0ex}
        {\noindent\bf Proof sketch:\ }}{\eop\vspace{1.5ex}}

\renewcommand{\texttt}[1]{{\small\textsf{#1}}}

\newcommand{\aff}{\kw{AFF}}

\newcommand{\dist}{\kw{dist}}
\newcommand{\hop}{\kw{hop}}

\newcommand{\diameter}[1]{\kw{dia}_{#1}}
\newcommand{\density}[1]{\kw{den}_{#1}}

\newcommand{\eps}{\prec}

\newcommand{\Reps}{M}

\newcommand{\eeps}{\lhd_r}

\newcommand{\ball}[1]{\hat{G}{#1}}

\newcommand{\algorithm}[1]{$\mathcal{#1}$}

\newcommand{\teamF}[1]{k{\sc TF${#1}$}\xspace}
\newcommand{\dynteamF}[1]{k{\sc DTF${#1}$}\xspace}

\newcommand{\mindia}{\kw{minDia}}
\newcommand{\minsumdis}{\kw{minSum}}
\newcommand{\denalk}{\kw{denAlk}}

\newcommand{\citationd}{{\sc Citation}\xspace}
\newcommand{\synthetic}{{\sc Synthetic}\xspace}
\newcommand{\youtube}{{\sc YouTube}\xspace}

\RequirePackage{url}
\def\doi#1{\def\@doi{#1}}
\doi{10.1145/1235}
\def\printdoi#1{\url{#1}}

\newcommand{\eetitle}[1]{\vspace{0.75ex}\noindent{\em #1}}

\title{Graph Pattern Matching for Dynamic Team Formation}

\author{
\vspace{0.5ex}
Shuai Ma \hspace{2ex} Jia Li \hspace{2ex} Chunming Hu \hspace{2ex} Xudong Liu \hspace{2ex} Jinpeng Huai\\
\vspace{0.5ex}
{\affaddr SKLSDE Lab, Beihang University, China}\\
{\affaddr Beijing Advanced Innovation Center for Big Data and Brain Computing, Beijing, China
}\\
{\affaddr\{mashuai, lijia1108, hucm, liuxd, huaijp\}@buaa.edu.cn}
}

\date{}
\begin{document}

\maketitle

\begin{abstract}
Finding a list of $k$ teams of experts, referred to as {\em top-$k$ team formation}, with the required skills and high collaboration compatibility has been extensively studied.
However, existing methods have not considered the specific collaboration relationships among different team members, \ie structural constraints, which are typically needed in practice.
In this study, we first propose a novel graph pattern matching approach for top-$k$ team formation, which incorporates both structural constraints and capacity bounds.
Second, we formulate and study the {\em dynamic top-$k$ team formation} problem due to the growing need of a dynamic environment.
Third, we develop an unified incremental approach, together with an optimization technique, to handle continuous pattern and data updates, separately and simultaneously, which has not been explored before.
\eat{
To
which as far as we know, has never been considered by previous study.
Different with existing work on devising incremental algorithms for data updates that can only compute changes to results in response to updates, there are no incremental algorithms for pattern updates with performance guarantees due to the inherent hardness, which to our knowledge has not been investigated before.
}
Finally, using real-life and synthetic data, we conduct an extensive experimental study to show the effectiveness and efficiency of our graph pattern matching approach for (dynamic) top-$k$ team formation.
\end{abstract}

\section{Introduction}
\label{sec-intro}

The {\em top-$k$ team formation} problem is to find a list of $k$ highly collaborative teams of experts such that every team satisfies the skill requirements of a certain task.
Various approaches~\cite{Lappas09,Kargar11,ArisLuca12,GajewarS12,realTeamForm13,SamikKVM12} have been proposed,
and fall into two categories in terms of the way to improve the collaborative compatibility of team members:
 (a) minimizing team communication costs,
defined with \eg the diameter, minimum spanning tree and the sum of pairwise member distances of the induced subgraph~\cite{Lappas09,Kargar11,ArisLuca12,SamikKVM12}, and
(b) maximizing team communication relations,
\eg the density of the induced subgraph~\cite{GajewarS12,realTeamForm13}.
Further, \cite{GajewarS12} and \cite{realTeamForm13} consider a practical setting by introducing a lower bound on the number of individuals with a specific skill in a team, and an upper bound of the total team members, respectively.

\begin{figure}[tb!]
\vspace{-2ex}
\begin{center}
\includegraphics[scale=1.15]{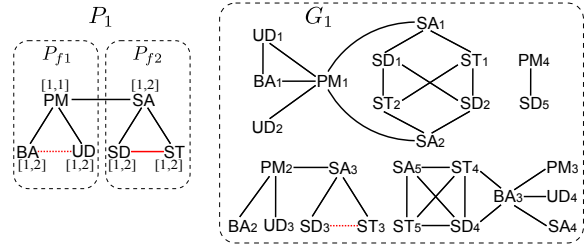}
\end{center}
\vspace{-3ex}
\caption{Motivation example}
\label{fig-motivation-example}
\vspace{-4ex}
\end{figure}

\begin{example}
\label{exm-motivation}
Consider a recommendation network $G_1$ taken from \cite{TerveenM05} as depicted in Fig.~\ref{fig-motivation-example},
in which (a) a node denotes a person labeled with her expertise, \eg project manager (\kw{PM}), software architect (\kw{SA}), software developer (\kw{SD}), software tester (\kw{ST}),
user interface designer (\kw{UD}) and business analyst (\kw{BA}),
and (b) an edge indicates the collaboration relationship between two persons, \eg ($\kw{PM_{1}}$, $\kw{UD_{1}}$) indicates $\kw{PM_{1}}$ worked well with $\kw{UD_{1}}$ within previous projects.

A headhunt helps to set up a team for a software product by searching proper candidates from $G_1$ (ignore dashed edges).
A desired team has
(i) one \kw{PM}, and one to two \kw{BAs}, \kw{UDs}, \kw{SAs}, \kw{SDs} and \kw{STs}, such that
(ii) \kw{PM} should collaborate with \kw{SAs}, \kw{BAs} and \kw{UDs} well, and \kw{SDs} and \kw{STs} should collaborate with each other well and both with \kw{SAs} well.

\eat{
One may want to look for candidate teams with existing methods,
 (1) by minimizing the {\em team diameter} \cite{Lappas09},
which returns the team with $\{\kw{BA_3}$, \kw{PM_3}, \kw{UD_4}, \kw{SA_4}, \kw{SD_4}, $\kw{ST_4}\}$,
(2) by minimizing the {\em sum of all-pair distances} of teams \cite{Kargar11},
which returns exactly the same team as (1) in this case,
or (3) by maximizing the {\em team density} \cite{GajewarS12}, which returns the team consisting of the nodes of the two connected components in $G_1$ with \kw{PM_1} and \kw{BA_3} (excluding \kw{UD_2}, \kw{PM_3}, \kw{UD_4}, \kw{SA_4}).

However, one may notice that these teams only satisfy the skill requirement, and cannot guarantee  the specific collaboration relationships among team members, \ie structural constraints. Indeed, the team found in (1) and (2) is connected by \kw{BA_3} only, and the team found in (3) has loose collaborations among its members.
Further, the capacity constraint in \cite{GajewarS12,realTeamForm13} needs to be extended to express the capacity bounds here.
That is, existing methods are not appropriate for identifying the above desired teams.
}

One may verify that existing methods \cite{Lappas09,Kargar11,GajewarS12,realTeamForm13}, can hardly find a desired team.
They only find teams satisfying the skill requirement~\cite{Lappas09,Kargar11,GajewarS12} and the lower bound capacity requirement~\cite{GajewarS12,realTeamForm13}  (condition (i)), and cannot guarantee  the specific collaboration relationships among team members, \ie structural constraints (condition (ii)).
 \end{example}

\eat{ It is because existing methods fall short of capturing the collaboration relations between specific team members, \ie the topology constraint (condition (2)).
However, in practice, it is rather important to take the functionality of different members into consideration~\cite{MerzSeeber15};
and few of them consider the capacity constraint (in condition (1)), and even consider, they always focus on the lower bounds~\cite{GajewarS12}, which results in a large quantity.

 and the number of team members is less than the cardinality requirement of the task
One can verify none of existing methods suffice to express the above requirements.

One can find none of the teams could work well in practice: the team found by (a) and (b) is connected by only one member \kw{BA} and the number of team members is less than the cardinality requirement of the task;
and the size of the team found by (c) is too large. It is because existing methods fall short of capturing the collaboration relations between specific team members, \ie the topology constraint (condition (2)).
However, in practice, it is rather important to take the functionality of different members into consideration~\cite{MerzSeeber15};
and few of them consider the capacity constraint (in condition (1)), and even consider, they always focus on the lower bounds~\cite{GajewarS12}, which results in a large quantity.}

\eat{
A natural question is how to further capture the {\em structural and capacity constraints} in a unified model for team formation?
We essentially introduce a revision of graph pattern matching, an extension of graph simulation~\cite{infsimu95} and strong simulation~\cite{MaCFHW14}, for team formation to fill in this gap.
}

A natural question is how to further capture the {\em structural and capacity constraints} in a unified model for team formation?
We  introduce a revision of {\em graph pattern matching} for team formation to fill in this gap.
Given a pattern graph $P$ and a data graph $G$, graph pattern matching is to find all subgraphs in $G$ that match $P$, and has been extensively studied~\cite{Ullmann76,infsimu95,FanLMTWW10,MaCFHW14,Guanfeng15,FanCount16}.  Essentially, we utilize patterns to capture the structural constraint, and revise the semantics of graph pattern matching for team formation. For instance, a desired team requirement can be specified by the pattern $P_1$ (ignore dashed edges) in Fig.~\ref{fig-motivation-example}, in which nodes represent the skill requirements,  edges specify the topology constraint, and the bounds on nodes are the capacity constraint.


Another issue lies in that team formation is accompanied with a highly dynamic environment. It typically needs many efforts to find the ideal teams, and is common for professionals to refine patterns  (requirements) multiple rounds~\cite{SajjadPG12,HabibiP15}. Further, real-life graphs are often big and constantly evolve over time~\cite{FanWW13-tods}. We show this with an example.

\begin{example}
\label{exm-motivation-inc}
Consider $P_{1}$ and $G_{1}$ in Example~\ref{exm-motivation} again.
	
\sstab{(1)} 
One may find that $P_1$ is too restrictive to find any sensible match in  $G_{1}$.
Hence, she needs to refine the pattern by updating $P_1$ with $\Delta P_{1}$, \eg an edge
deletion $(\kw{SD},\kw{ST})^-$.
	
\sstab{(2)} It is also common that a data update $\Delta G_1$ comes on $G_1$, \eg an edge insertion $(\kw{SD_3},\kw{ST_3})^+$.
	
\sstab{(3)} Finally, it can be the case when pattern update $\Delta P_{1}$  and data update $\Delta G_{1}$ come simultaneously on $P_1$ and $G_1$.
\end{example}

This motivates us to study the {\em dynamic top-k team formation} problem, to handle continuous
pattern and data updates, separately and simultaneously. It is known that incremental algorithms avoid re-computing from scratch by re-using previous results~\cite{inc-survey}. However, incremental algorithms of graph pattern matching for pattern updates has not been investigated, though there exist incremental algorithms of graph pattern matching for dealing with data updates~\cite{FanLMTWW10,FanWW13-tods,FanHT17}.
Further, it is also challenging for incremental algorithms to handle simultaneous pattern and data updates  in a unified way.

\eat{
often too costly to recompute top-$k$ teams from scratch in response to pattern and data updates.
These highlight the need for incremental algorithms to compute the updated top-$k$ teams.
As opposed to batch algorithms, incremental algorithms avoid re-computing from scratch by re-using previous results.
Traditional incremental algorithms for graph queries including graph pattern matching queries are proposed to compute changes in response to data updates, which has a lot of work~\cite{FanLMTWW10,FanWW13-tods,FanHT17}.
However, the study on incremental algorithms for graph pattern updates is in its vacancy.

Given a pattern graph $P$ and a data graph $G$, the graph pattern matching results $P(G)$ in $G$ for $P$ and changes $\Delta G$ to $G$ as input,
the traditional incremental matching problem is to compute changes $\Delta O$ to $P(G)$ such that $P(G\oplus\Delta G)$=$P(G)\oplus \Delta O$.
Here $\oplus$ denotes applying changes $\Delta S$ to $S$, when $S$ is data graph $G$ or query result $P(G)$.
It is to improve response time by reducing computations on big $G$ to small $\Delta G$ and $\Delta O$.

As a basis of \dynteamF, we also investigate the incremental problem for graph pattern matching on pattern updates.
The definition can be formulated along the same line with existing data incremental problems.
However, devising pattern incremental algorithms is quite challenging as the impact of pattern updates is global,
due to the inherent hardness of the problem.
Despite of this, we devise an effective unified incremental approach for the \dynteamF\, problem.

 }

\looseness = -1

\stitle{Contributions}. To this end,  we introduce a graph pattern matching approach for (dynamic) top-$k$ team formation.

\stab (1) We propose {\em team simulation}, a revision of traditional graph pattern matching, for top-$k$ team formation (Section~\ref{sec-tsim}).
It extends existing methods by incorporating the structural and capacity constraints using pattern graphs.
To cope with the highly dynamic environment of team formation, we also formulate the dynamic top-$k$ team formation problem  (Section~\ref{sec-tsim}), for dealing with pattern and data updates, separately and simultaneously.

\stab (2) We develop a batch algorithm for computing top-$k$ teams via team simulation (Section~\ref{sec-tsimAlg}).
We study the satisfiability problem for pattern graphs, a new problem raised in the presence of capacity bounds for graph pattern matching.
We also propose two optimization techniques, handling radius varied balls and density based filtering, for speeding up the process of computations.

\stab (3) We develop a unified approach to handling the need for both pattern and data updates (Sections~\ref{sec-dynamictopk} and~\ref{sec-IncAlg}).
Due to the inherent difficulty of the problem, we propose an incremental strategy based on {\em pattern fragmentation} and {\em affected balls} by localizing the effects of pattern and data updates, and we develop a unified incremental algorithm for dealing with separate and simultaneous pattern and data updates,
with an optimization technique with the {\em early return} property for incremental top-$k$ algorithms, an analogy of the traditional early termination property.

\eat{

As opposed to batch algorithms that recompute new match results from scratch,
incremental algorithms \cite{inc-survey} have been proposed to compute new match results,
by reusing and updating previous cached results according to the changes in input.
\looseness=-1

\begin{example}
\label{exm-motivation-inc}
Recall $P_{1}$ and $G_{1}$ in Fig.~\ref{fig-motivation-example} and continue.

\sstab{(1)} When the user gets the matched teams $P_{1}(G_{1})$ to $P_{1}$ in $G_{1}$, she may find $P_1$ is too restrictive to find possible teams.
Therefore, she enters pattern updates $\Delta P_1$ on $P_1$, composed of an edge deletion $(\kw{SD},\kw{ST})^-$.
We denote $P_{1}\oplus\Delta P_{1}$ the revised pattern.
It is much better to derive $P_{1}\oplus\Delta P_{1}(G_{1})$ from $P_{1}(G_{1})$
than a computation for $P_{1}\oplus\Delta P_{1}$ in $G_{1}$ from scratch.
\looseness=-1

\sstab{(2)} When data updates $\Delta G_1$ come on $G_1$, composed of an edge insertion $(\kw{SD_3},\kw{ST_3})^+$.
Similarly with (1), it is better to compute $P_{1}(G_{1}\oplus\Delta G_{1})$ from $P_{1}(G_{1})$ incrementally.

\sstab{(3)} Even better, when it comes to the case pattern $\Delta P_{1}$ and data updates $\Delta G_{1}$ come together, it is desirable to compute $P_{1}\oplus\Delta P_{1}(G_{1}\oplus\Delta G_{1})$ from $P_{1}(G_{1})$ incrementally.
\end{example}
}

\eat{
These encourage us to propose the {\em dynamic top-$k$ team formation} problem (\dynteamF),
to handle separate and simultaneous pattern and data updates.
However, the problem is non-trivial.
There has been a bunch of work on data incremental algorithms on graphs~\cite{Reps96,FanWW13-tods},
but less work on handling pattern updates is known, and furthermore less work on handling both data and pattern updates is known. The challenges comes from three aspects:
(a) the nature of pattern updates, pattern updates have significant impacts on the match results.
A unit update is likely to result in the entire change in match results, such that all previous match results need to be re-computed, which is actually a costly batch computation;
(b) the demand to design light weight and effective auxiliary data structures for the dynamic team formation problem, concerning the space consumption and the efficiency;
and (c) the requirement to design a {\em one-method-fits-two} model, to support pattern and data updates simultaneously and separately.

To solve the {\em top-$k$ team formation} problem and the {\em dynamic top-$k$ team formation} problem, several fundamental problems call for a full treatment.
(1) How to extend graph pattern matching for top-$k$ team formation to retain the properties in existing methods, and pursue more practical usage,
without sacrificing the efficiency?
(2) How to efficiently support pattern updates for, \eg user query adjustments on patterns, by reusing previous cached results as much as possible?
(3) How to handle data updates to meet the practical requirements?
(4) How to design a unified incremental model to handle both pattern and data updates efficiently in a simultaneous manner?
}

\eat{
\stitle{Contributions}. To this end,  we introduce a graph pattern matching approach for (dynamic) team formation.

\stab (1) We propose {\em team simulation} for top-$k$ team formation, a revision of traditional graph pattern matching,
which extends existing methods {\em strong simulation} for team formation by incorporating the structural and capacity constraints using pattern graphs  (Section~\ref{sec-tsim}).
We also study the satisfiability problem for pattern graphs, a new problem raised in the presence of capacity bounds for graph pattern matching.
We finally develop an efficient cubic batch algorithm with two optimization techniques, \ie density based filtering and pattern minimization, for computing top-$k$ teams via team simulation.

\stab (2) We formulate the dynamic top-$k$ team formation problem, and develop a unified approach to handling the need for both pattern and data updates (Sections~\ref{sec-dynamictopk} \& \ref{sec-IncAlg}).
We first prove that the  problem is unbounded even for single pattern or data updates, by extending the notion of incremental boundedness~\cite{Reps96}.
In light of this, we then propose an incremental strategy based on {\em pattern fragmentation} and {\em affected balls} to localize the effects of pattern and data updates. We finally develop a unified incremental algorithm for dealing with continuous pattern and data updates, separately and simultaneously,
with an optimization technique with the {\em early return} property for incremental top-$k$ algorithms, an analogy of the traditional early termination property.
}

\eat{
\stab (3) Using real-life and synthetic graphs, we experimentally verify the effectiveness and efficiency of the static and dynamic top-$k$ graph pattern matching model and optimization techniques, for separate and simultaneous pattern and data updates, for solving \teamF\,and \dynteamF\,problems.
{\bf Add details}
}

\stab (4) Using real-life data (\citationd) and synthetic data (\synthetic), we demonstrate the effectiveness and efficiency of our graph pattern matching approach for (dynamic) team formation (Section~\ref{sec-expt}).
We find that (a) our method is able to identify more sensible teams than existing team formation methods \wrt \eat{three} practical measurements,
and (b) our incremental algorithm outperforms our batch algorithm, even when changes reach 36\% for pattern updates, 34\% for data updates and (25\%, 22\%) for simultaneous pattern and data updates, and when 29\% for continuous pattern updates, 26\% for continuous data updates and (20\%, 18\%) for continuously simultaneous pattern and data updates, respectively.

To our knowledge, this work is among the first to study simultaneous pattern and data incremental computations, no previous work has studied pattern updates for incremental pattern matching~\cite{FanHT17,FanWW13-tods}, not to mention continuous and simultaneous pattern and data updates.  This is  the most general dynamic setting for incremental computations.

All detailed proofs are available  in the full version~\cite{fullvldb18}.

\stitle{Related work}. Previous work can be classified as follows.


 Graph simulation \cite{infsimu95} and its extensions have been introduced for graph pattern matching~\cite{FanLMTWW10,MaCFHW14,Guanfeng15,FanCount16}, in which {\em strong simulation} introduces duality and locality into simulation~\cite{MaCFHW14}, and shows a good balance between its computational complexity and its ability to preserve graph topology. Furthermore, \cite{FanCount16} already adopts capacity bounds on the edges of pattern graphs via subgraph isomorphism, and \cite{FanWWXin13} uses graph pattern matching to find single experts, instead of a team of experts. In this study, team simulation is proposed for team formation as an extension of graph simulation and strong simulation on undirected graphs with capacity constraints on the nodes of pattern graphs.


There has been a host of work on team formation by minimizing the communication cost of team members, based on the diameter, density, minimum spanning tree, Steiner tree, and sum of pairwise member distances among others \cite{Lappas09,Kargar11,ArisLuca12,GajewarS12,realTeamForm13,SamikKVM12,LiTongCao15}, which are essentially  a specialized class of keyword search on graphs~\cite{Aggarwal10}. Similar to~\cite{Kargar11}, we are to find top-$k$ teams. However, \cite{Kargar11} adopted Lawler's procedure \cite{Lawler1972}, and is inappropriate for large graphs. We also adopt {\em density}  as the communication cost, which shows a better performance~\cite{GajewarS12}, and further require that all team members are {\em close to each other} (located in the same balls), along the same line with~\cite{Lappas09,Kargar11,ArisLuca12,SamikKVM12}.
Except for simply minimizing the communication cost among team members,
\cite{HuangLv16,Kargar11} consider minimizing the cost among team members and team leaders.
Different from these work, we introduce {\em structural constraints}, in terms of graph pattern matching~\cite{FanLMTWW10,MaCFHW14}, into team formation, while retaining the capacity bounds on specific team members like~\cite{realTeamForm13,GajewarS12}.


Incremental algorithms (see \cite{inc-survey,FanHT17} for a survey) have proven usefulness in a variety of applications,
and have been studied for graph pattern matching~\cite{FanLMTWW10,FanWW13-tods} and team formation \cite{ArisLuca12} as well.
However, \cite{inc-survey,FanHT17,FanLMTWW10,FanWW13-tods} only consider data updates,
and \cite{ArisLuca12} only considers continuously coming new tasks. In this work, we  deal with both pattern and data updates for team formation, and support both insertions and deletions. To our knowledge, this is the first study on pattern updates, and is the most general and practical dynamic setting considered so far.

{Query reformulation (\aka query rewriting) is to generate alternative queries that may produce better answers,
and has been studied for structured queries~\cite{MottinEmpty13}, keyword queries~\cite{YaoQReform12} and graph queries~\cite{MottinGReform15}.
However, different from our study of handling pattern updates, the focus of query reformulation is not on incremental computations.


Although top-$k$ queries (see~\cite{IlyasBS08} for a survey) have been studied for both graph pattern matching and team formation~\cite{Kargar11},
they have never been studied for both team formation and graph pattern matching  in a dynamic setting.

\eat{
\etitle{Query reformulation}. \textcolor{blue}{Query reformulation (\aka query rewriting/modification) is to provide users with a set of specializations of the original query that may better capture their search intent.
It has been studied for structured queries~\cite{MottinEmpty13}, keyword queries~\cite{YaoQReform12} and graph queries~\cite{MottinGReform15}.
However, they all aim at automatically providing alternative queries (\eg supergraphs~\cite{MottinGReform15}) to users,
while care less about answering the changed queries.
In this work, we focus on handling query modification from users and how to quickly return the answers based on previous answers.}
}

\eat{
To retain the properties in existing methods, and pursue more practical usage, we incorporate the capacity bounds, density measure and locality into graph simulation as the matching semantics.

When a user inputs a pattern graph into a system with a data graph already stored in it, and gets the match result, sometimes he or she may find that the query is not appropriate in some sense after an analysis on the result. Then the user probably makes some small adjustments on the pattern graph. However, it is expensive to recompute the matches from scratch via batch algorithm due to the slight change in pattern graph, and this comes the convenience for executing incremental graph matching, which just computes changes in the match result corresponding to the updates in pattern graph.
}

\newcommand{\grouprec}{\kw{batch}}
\newcommand{\optgrouprec}{\kw{batch}}
\newcommand{\minp}{\kw{minP}}
\newcommand{\rgraphsim}{\kw{undirgSim}}
\newcommand{\incsim}{\kw{incSim}}

\section{Dynamic Team Formation}
\label{sec-tsim}

We first propose {\em team simulation}, a revision of traditional graph pattern matching.
We then formally introduce the {\em top-$k$ team formation} problem via team simulation.
We finally present the {\em dynamic top-$k$ team formation} problem.

\eat{
In this section, we first present basic notations of graphs, and then propose team simulation, an extension of traditional graph pattern matching.
We finally formally introduce the top-$k$ team formation problem via team simulation, and propose a cubic algorithm with two optimization techniques.

\subsection{Data Graphs and Pattern Graphs}
\label{subsec-graphdefs}

\stitle{Data graphs}.
A {\em data graph} is a labeled undirected graph $G(V, E, l)$, where
(1) $V$ is a finite set of nodes;
(2) $E \subseteq V \times V$, in which $(u, u')$ or $(u', u)$ $\in E$ denotes an undirected edge between nodes $u$ and $u'$; and
(3) $l$ is a total labeling function that maps each node in $V$ to a set of labels.


Intuitively, the node labels carry the content of the node, \eg skills, keywords, blogs, rating~\cite{AmerYahiaBB07}.
The edges specify the affinity or collaborative compatibility among nodes~\cite{GajewarS12}.

\stitle{Pattern graphs}.
A {\em pattern graph} (or simply pattern) is an undirected graph $P(V_P$, $E_P$, $l_P$, $f_P)$, in which
(1) $V_P$ and $E_P$ are the set of nodes and the set of edges, respectively;
(2) $l_P$ is a total labeling function that maps each node in $V_P$ to a single label; and
(3) $f_P$ is a total capacity function such that for each node $u\in V_P$, $f_P(u)$ is a closed interval $[x, y]$, where $x\le y$ are non-negative integers.

Intuitively, $f_P(u)$ specifies a range bound for node $u$,
indicating the required quantity for the matched nodes in data graphs.
Note that for traditional graph patterns~\cite{Galla06, b-matching,FanLMTWW10,FanCount16}, bounds are typically carried on edges, not on nodes.
\looseness=-1

\begin{example}
\label{exm-pattern}
Consider pattern graph $P_1$ and data graph $G_1$ in Fig.~\ref{exm-motivation}. Observe that there is a capacity bound on each node in $P_1$.
For example, capacity bound [1,2] on node \kw{SD} in $P_1$ indicates in the match result,
there should be at least one and at most two nodes matched with \kw{SD}.
\end{example}

\stitle{Remarks}. We assume \kwlog~that pattern graphs are connected, as a common practice.
We also simply denote data graphs $G(V, E, l$) as  $G(V, E)$ and pattern graphs $P(V_P$, $E_P$, $l_P$, $f_P)$ as $P(V_P$, $E_P)$, respectively.
Furthermore, the size $|G|$ of data graphs $G$ (resp. $|P|$ of pattern graphs $P$) is the total number of nodes and edges in $G$ (resp. $P$).
}

\eat{
We shall use the following notation
	
\etitle{Subgraphs}.
Graph $H(V_s, E_s,  l_{H})$ is a {\em subgraph} of data graph $G(V, E, l)$, denoted as $G[V_s, E_s]$, if
(1) for each node $u\in V_s$, $u\in V$ and $l_{H}(u) = l(u)$, and
(2) for each edge $e\in E_s$, $e\in E$. 
That is, subgraph $G[V_s, E_s]$ only contains a subset of nodes and a subset of edges of graph $G$.

We also denote subgraph $G[V_s, E_s]$ as subgraph $G[V_s]$, if $E_{s}$ is exactly

Similarly, we can define {\em subgraphs for pattern graphs}, which further requires the preservation of the node capacity.

\etitle{Induced subgraphs}. An {\em induced subgraph} $G_{s}(V_{s}, E_{s})$ is composed of vertices $V_{s}$ who is a subset of the vertices $V$ of graph $G(V, E)$, and together with the edge set $E_{s}$ whose endpoints are both in $V_{s}$ in $G(V, E)$.

\etitle{Subpatterns}.
Pattern $Q(V_s, E_s,  l_{q}, f_{q})$ is a {\em subpattern} of pattern $P(V_p$, $E_p$, $l_p$, $f_p)$, denoted as $P[V_s, E_s]$, if
(1) for each node $u\in V_s$, $u\in V$, $l_{q}(u) = l_p(u)$ and $f_{q}(u) = f_p(u)$, and
(2) for each edge $e\in E_s$, $e\in E$. 
That is, subpattern $G[V_s, E_s]$ only contains a subset of nodes and a subset of edges of pattern $P$.}


\eat{
\etitle{Neighbors}. We say that node $u'$ is a {\em neighbor} of node $u$ if there is an edge between $u$ and $u'$ in a data or pattern graph.

\etitle{Paths}.
A {\em simple path} (or simply a {\em path}) $\rho$ is a sequence of nodes $v_1/\ldots/v_n$ with no repeated nodes, and, moreover, for each $i\in[1, n-1]$, $(v_i$, $v_{i+1})$ is an edge in $G$. The {\em length} of a path $\rho$ is the number of edges in $\rho$.

A graph is {\em connected} if for each pair of nodes, there exists a  path connecting them.

\etitle{Distances}. Given two nodes $v$ and $v'$ in a graph $G$, the {\em  distance} from $v$ to $v'$,
denoted by $\dist(v,v')$, is  the shortest length of all {\em paths}
from $v$ to $v'$ in $G$.
}

\eat{
\etitle{Diameter}. The {\em diameter} of a connected graph $G$,  denoted by $\diameter{G}$, is the longest shortest distance of all pairs of nodes in $G$, \ie $\diameter{G}$ = $\kw{max}(\kw{dis}(v,v'))$ for all nodes $v,v'$ in $G$.
}

\eat{
\etitle{Distances}. The {\em length} of a path $\rho$ is
the sum of the weights of its constituent edges, \ie $\sum_{i=1}^{n-1} w(v_i, v_{i+1})$.

Given two nodes $v, v'$ in a graph $G$, the {\em distance} from $v$ to $v'$,
denoted by $\dist(v,v')$, is the length of the shortest {\em path}
from $v$ to $v'$ in $G$.
}

\subsection{Team Simulation}
\label{subsec-extsim}

We first extend pattern graphs of traditional graph pattern matching to carry capacity requirements,
and then define team simulation on undirected graphs.

We start with basic notations.

\stitle{Data graphs.} A {\em data graph} is a labeled undirected graph $G(V$, $E$, $l)$, where $V$ and $E$ are the sets of nodes and edges, respectively; and $l$ is a total labeling function that maps each node in $V$ to a set of labels.

\stitle{Pattern graphs}.
A {\em pattern graph} (or simply pattern) is an undirected graph $P(V_P$, $E_P$, $l_P$, $f_P)$, in which
(1) $V_P$ and $E_P$ are the set of nodes and the set of edges, respectively;
(2) $l_P$ is a total labeling function that maps each node in $V_P$ to a single label; and
(3) $f_P$ is a total capacity function such that for each node $u\in V_P$, $f_P(u)$ is a closed interval $[x, y]$, where $x\le y$ are non-negative integers.

Intuitively, $f_P(u)$ specifies a range bound for node $u$,
indicating the required quantity for the matched nodes in data graphs.
Note that for traditional patterns~\cite{Galla06, b-matching,FanLMTWW10,FanCount16}, bounds are typically carried on edges, not on nodes.
We also also denote data and pattern graphs as $G(V$, $E)$ and $P(V_{P}$, $E_{P})$ respectively. The size of $G$ (resp. $P$), denoted by $|G|$ (resp. $|P|$), is defined to be the total number of nodes and edges in $G$ (resp. $P$).


\eat{As strong simulation is an extension of graph simulation, which is also defined on directed graphs~\cite{infsimu95,FanLMTWW10},
We now redefine graph simulation on undirected graphs. Consider a pattern graph $P(V_P$, $E_P)$ and a data graph $G(V$, $E)$.
}

We now redefine graph simulation on undirected graphs, which is originally defined on directed graphs~\cite{infsimu95,FanLMTWW10}. Consider pattern graph $P(V_P$, $E_P)$ and data graph $G(V$, $E)$.

\stitle{Graph simulation}. Data graph $G$ {\em matches} pattern graph $P$ via
graph simulation, denoted by $P \eps G$, if there exists a binary {\em match relation} $\Reps \subseteq V_P \times V$ in $G$ for $P$ such that

\vspace{0.5ex}
\ni
(1) for each $(u, v) \in \Reps$, the label of $u$ matches one label in the label set of $v$, \ie $l_{P}(u) \in l(v)$; and

\vspace{0.5ex}
\ni
(2) for each node  $u\in V_P$, there exists $v\in V$ such that
(a) $(u, v) \in \Reps$, and
(b) for each adjacent node $u'$ of $u$ in $P$, there exists a adjacent node $v'$ of $v$
in $G$ such that $(u',v') \in \Reps$.

For any $G$ that matches $P$, there exists a {\em unique maximum} match relation via graph simulation~\cite{infsimu95}.

We then introduce the notions of balls and match graphs.


\eat{
Intuitively, graph simulation preserves the label match and
neighborhood relationships between pattern graphs and the matched data graphs~\cite{FanLMTWW10,MaCFHW14}.

Following from \cite{FanLMTWW10,MaCFHW14}, the revised graph simulation above is well-defined, as shown below.

\begin{prop}
\label{prop-sim-maximum-match}
For any data graph $G$ and pattern graph $P$  such that $P\eps G$, via graph simulation, there is a unique maximum match relation in $G$ for $P$.
\end{prop}
}

\etitle{Balls}. For a node $v$ in data graph $G$ and a non-negative integer $r$,
the {\em ball} with {\em center} $v$ and {\em radius} $r$  is a subgraph of $G$,
denoted by $\ball{[v, r]}$, such that (1) all nodes $v'$ are in $\ball{[v, r]}$, if
the number of hops between $v'$ and $v$, $\hop(v', v)$, is no more than $r$, and (2) it has exactly the edges
appearing in $G$ over the same node set.

\eat{
Intuitively, a ball is a connected graph such that all node pairs have bounded hops.
Indeed, as observed in~\cite{Buchan2004}, when social distance increases, the
closeness of relationships decreases and the  relationships may become
irrelevant. Hence it often suffices in practice to consider only those
matches of a pattern graph that fall in a small ball.
}

\etitle{Match graphs}. The {\em match} graph $\wrt$ a binary relation $\Reps\subseteq V_P\times V$ is a subgraph $G_s$ of data graph $G$, in which
(1)  a node $v\in V_s$ if and only if it is in $\Reps$, and
(2) it has exactly the edges
appearing in $G$ over the same node set.

Intuitively, the match graph $G_s$ $\wrt$ $\Reps$ is the induced subgraph of
$G$ such that its nodes play a role in $\Reps$.

\eat{
\stitle{Induced match graphs}. Consider a binary relation $\Reps\subseteq V_q\times V$.
The {\em induced match} graph $\wrt$ $\Reps$ is a induced subgraph $G[V_s, E_s]$ of $G$, in which
(1) a node $v\in V_s$ if and only if it is in $\Reps$, and
(2) an edge $(v,v')\in E_s$ if and only if $v$ and $v'$ are in $\Reps$.
Intuitively, the induced match graph $G[V_s, E_s]$ $\wrt$ $\Reps$ is the induced subgraph of $G$ such that each of its nodes plays a role in $\Reps$, together with the adjacent edges in $G$.
}

We are now ready to define team simulation, by extending graph simulation to incorporate the locality constraints enforced by balls, and the capacity bounds carried by patterns.
\looseness=-1

\stitle{Team simulation}. Data graph $G$ {\em matches} pattern $P$  via
team simulation \wrt a radius $r$, denoted by $P \eeps G$, if
there exists a {\em ball} $\ball{[v, t]}$ ($t \in [1,r]$, $t \in Z$) in $G$, such that

\vspace{0.5ex}
\ni(1) $P\eps \ball{[v, t]}$, with the maximum match relation $\Reps$ and the match graph $G_s$ $\wrt$ $\Reps$; and

\vspace{0.5ex}
\ni(2) for each node $u$ in $P$, the number of nodes $v$ in $G_s$  with $(u, v)\in \Reps$ falls into $f_P(u)$.

\vspace{0.5ex}
We refer to  $G_s$ as a {\em perfect} subgraph of $G$ \wrt $P$.

Intuitively, (1) pattern graphs $P$ capture the structural and capacity constraints, and (2) a perfect subgraph $G_s$ of pattern $P$ corresponds to a desired team, which is required to satisfy the following conditions:
(a) $G_s$ itself is located in a ball $\ball{[v, t]}$  where $t \in [1,r]$ as a match graph; and
(b) $G_s$ satisfies the capacity constraints carried over pattern $P$.

\begin{example}
\label{exm-rsimulation}
Consider pattern $P_1$ and data graph $G_1$ in Fig.~\ref{exm-motivation}, and  team simulation with $r=2$ is adopted.

One can easily verify that $P_1$ matches $G_1$  via team simulation, \ie $P_1 \eeps G_1$,
as (a) there is a perfect subgraph in in ball
$\ball{[\kw{PM_1}, 2]}$, \ie the connected component of $G_1$ containing $\kw{PM_{1}}$, which maps \kw{PM}, \kw{BA}, \kw{UD}, \kw{SA}, \kw{SD} and \kw{ST} in $P_1$ to \kw{PM_1}, \kw{BA_1}, \{\kw{UD_1}, \kw{UD_2}\}, \{\kw{SA_1}, \kw{SA_2}\}, \{\kw{SD_1}, \kw{SD_2}\} and \{\kw{ST_1}, \kw{ST_2}\}, respectively, and, moreover, (b) the capacity bounds on all pattern nodes are satisfied.
\end{example}

\eat{
\stitle{Remarks}.
(1) Graph simulation is a special case of team simulation on undirected graphs,
when the capacity on each pattern node is $[1, +\infty]$, and $r$ is no less than the diameter of data graphs.
(2) Further, strong simulation is also a special case of team simulation on undirected graphs,  when the capacity on each pattern node is $[1, +\infty]$,
$r$ is equal to the diameter of pattern graphs, and match graphs only involve with those edges matching edges in pattern graphs.
}

\vspace{-1ex}
\stitle{Remarks}.
(1) Team simulation differs from graph simulation~\cite{infsimu95} and strong simulation~\cite{MaCFHW14} in the existence of capacity bounds on pattern graphs and its ability to capture matches on undirected graphs.

\sstab(2) Different from strong simulation with a fixed radius for balls (\ie the diameter of a pattern), team simulation adopts a more natural setting that the radius of balls is auto-adjustable, having a user specified upper bound only.


\eat{
\stitle{Remarks}.
(1) Graph simulation is a special case of (1) team simulation for undirected graphs,
when the capacity on each pattern node is $[1, +\infty]$ and $r$ is no less than the diameter of data graphs.
(2) Graph simulation~\cite{FanLMTWW10} and dual simulation~\cite{MaCFHW14} are equal on undirected graphs.
Hence, many good properties of dual simulation naturally carry over to graph simulation on undirected graphs.
(3) {\bf The difference of matched graphs between strong simulation and T-simulation.}
}



\eat{

When given a pattern graph, we first make a examination on it ensuring that there exist data graphs in which we can find matches using the pattern graph. If the answer is true, we can execute the following computing work. Otherwise, we require the user to retype in another pattern graph.
}

\eat{
\begin{figure}[tb!]
\begin{center}
\includegraphics[scale=1.28]{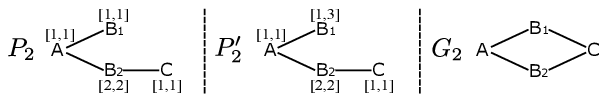}
\end{center}
\vspace{-3ex}
\caption{Pattern graphs}
\vspace{-4ex}
\label{fig-consistency-example}
\end{figure}

\stitle{Pattern satisfiability}.
We say that a pattern $P$ is {\em satisfiable} iff there exists a data graph $G$ such that $P \eeps G$.

Different from graph simulation \cite{infsimu95} and its extensions \cite{FanLMTWW10,MaCFHW14},
pattern graphs may be unsatisfiable for team simulation, due to the presence of capacity constraints enforced on patterns. We illustrate this with an example below.

\begin{example}
\label{exm-consistency}
\ni(1)  Consider pattern $P_2$ in Fig.~\ref{fig-consistency-example}.
One can verify that there exist no data graphs $G$ such that $P_2 \eeps G$ because (a) for any nodes $u$ in $G$, if $u$ matches with the node labeled with $B_2$, then it must match with the node labeled with $B_1$, and, hence, (b) the capacity upper bound on $B_1$ should not be less than the lower bound on $B_2$.

\sstab(2) Pattern $P_1$ in Fig.~\ref{exm-motivation} is satisfiable as $P_1 \eeps G_1$.
\end{example}

The good news is that checking the satisfiability of pattern graphs can be done in low polynomial time.

\begin{prop}
\label{prop-pattern-consistency}
The satisfiability of patterns $P$ can be checked in $O(|P|^2)$ time.
\end{prop}

\proofS By treating $P$ as both data and pattern graphs, compute the maximum match relation $M$ in $P$ for $P$, via graph simulation.
Indeed, pattern $P$ is satisfiable iff for each $(u, v)\in M$ with the capacity bounds $[x_u, y_u]$ on $u$ and $[x_v, y_v]$ on $v$, respectively, $x_v \leq y_u$ holds. Observe that the size of $M$ is bounded by $|P|^2$.
\eop

Pattern graphs are typically small, and we only consider  satisfiable patterns in the sequel, by Proposition~\ref{prop-pattern-consistency}.
}

\subsection{Top-k Team Formation}
\label{subsec-teamF}

 Given pattern $P$, data graph $G$, and two positive integers $r$ and $k$, the {\em top-$k$ team formation} problem, denoted as \teamF{(P, G, k)},
is to find a list $L_{k}$ of $k$ perfect subgraphs (\ie teams) with the top-$k$ largest density in $G$ for $P$, via team simulation.
\looseness=-1

Here the {\em density}\, $\density{G}$ of graph $G(V, E)$ is $|E|/|V|$, where $|E|$ and $|V|$ are the number of edges and the number of nodes respectively, as commonly used in data mining applications~\cite{maximumDenseSubgraph,EVMK12}.
Intuitively, the larger $\density{G}$ is, the more collaborative a team is.
In this way, not only the two objective functions of existing team formation methods are preserved,
\ie the locality retained by balls and the density function in selecting top-$k$ results,
but also the relationships among members and the capacity constraint on patterns.

\begin{example}
\label{exa-teamF}
Consider $P_1, G_1$ in Fig.~\ref{exm-motivation} and $r=2$.
We simply set $k=1$, as most existing solutions for \teamF{} only compute the best team \cite{Lappas09,ArisLuca12,GajewarS12,realTeamForm13,SamikKVM12}.

One may want to look for candidate teams with existing methods, satisfying the search requirement in Example \ref{exm-motivation}:
\ni(1) by minimizing the {\em team diameter} \cite{Lappas09},
which returns the team with $\{\kw{BA_3}$, \kw{PM_3}, \kw{UD_4}, \kw{SA_4}, \kw{SD_4}, $\kw{ST_4}\}$,

\ni(2) by minimizing the {\em sum of all-pair distances} of teams \cite{Kargar11},
which returns exactly the same team as (1) in this case, or

\ni(3) by maximizing the {\em team density} \cite{GajewarS12}, which returns the team with all the nodes in the two connected components in $G_1$ with \kw{PM_1} and \kw{BA_3}, except \kw{UD_2}, \kw{PM_3}, \kw{UD_4}, \kw{SA_4}.

One may already notice that these teams only satisfy the skill requirement, \ie condition (i) in Example \ref{exm-motivation}, and cannot guarantee  the specific collaboration relationships among team members.
Indeed, the team found in (1) and (2) is connected by \kw{BA_3} only, and the team found in (3) has loose collaborations among its members.
That is, existing methods are not appropriate for identifying the the desired teams.

When team simulation is adopted, it returns the perfect subgraph in Example \ref{exm-rsimulation} with its density = 1.4,
satisfying both conditions (i) and (ii), much better than those teams found by the above existing methods.
\end{example}

\eat{
\sstab
(a) Algorithm \mindia~\cite{Lappas09} is to find a team with the {\em minimum diameter}, and returns $\{\kw{BA_3}$, \kw{PM_3}, \kw{UD_4}, \kw{SA_4}, \kw{SD_4}, $\kw{ST_4}\}$.

\sstab
(b) Algorithm \minsumdis~\cite{Kargar11} aims at minimizing the sum of all-pair distances, which is an extension of \mindia, and it returns exactly the same team as \mindia in this case.

\sstab
(c) Algorithm \denalk~\cite{GajewarS12} is to find a team with the {\em largest density}, and
returns the subgraph consisting of the two connected components of $G_1$ containing \kw{PM_1} and \kw{BA_3} respectively (excluding \kw{UD_1},\kw{PM_3},\kw{UD_4},\kw{SA_4}), whose density is 1.5.

\sstab
(d) Our team simulation returns the perfect subgraph in Example \ref{exm-rsimulation} with its density = 1.4 as the best team.

One can see that only team simulation captures the topology, cardinality and collaboration requirements of teams.
}

\eat{
\begin{figure}[t!]
\begin{center}
{\small
\myhrule
\vspace{-3ex}
\mat{0ex}{
\sstab {\sl Input:\/} $G(V$, $E)$, $P(V_P$, $E_P)$, and positive integers $k, r$.\\
{\sl Output:\/} The list $L_{k}$ of top-$k$ teams\\
\sstab \bcc \hspace{1ex} \= $L_{k} := \emptyset$;\\
\icc \> \For each ball $\ball{[v,r]}$ in $G$ \Do\\
\icc \>\hspace{2ex}\= $G_s$ := \rgraphsim$(P, \ball{[v,r]})$;\\
\icc \>\> \If $\density{G_s}>$ the density of the $k$-th result in $L_{k}$ \Then\\
\icc \>\>\hspace{2ex}\= remove the $k$-th result in $L_{k}$ and insert $G_s$ into $L_{k}$;\\
\icc \> \Return $L_{k}$.
}

\vspace{-4ex} \myhrule
}
\end{center}
\vspace{-3ex}
\caption{Algorithm \grouprec} \label{alg-grouprec}
\vspace{-4ex}
\end{figure}

We now present an algorithm for \teamF{} via team simulation.

\stitle{Algorithm for Top-k Team Formation.} Given $P$, $G$, and two integers $r$ and $k$, an algorithm, referred to as \grouprec, is shown in Fig.~\ref{alg-grouprec}.
\grouprec is to find the set of perfect subgraphs $G_s$ by inspecting those balls $\ball{[v, r]}$ centered at each node $v$ of $G$, and returns the top-$k$ densest ones.
It computes the perfect subgraph $G_s$ of $P$ in the ball $\ball{[v,r]}$ via team simulation by invoking \rgraphsim(lines 2-3).
Procedure \rgraphsim is deferred to the appendix, which is an adaption from graph simulation~\cite{infsimu95,FanLMTWW10},
by extending from directed to undirected graphs,  incorporating the capacity constraint check, and only maintaining the capacity bounds satisfied perfect subgraphs.
In the process, \grouprec maintains the top-$k$ non-repeated perfect subgraphs with current top-$k$ largest density (lines 4-5), and returns them when all balls are processed (line 6).

\begin{example}
\label{exa-alg-batch} Consider $r=2$ and $k=2$.

\sstab(1) For $P_1$ and $G_1$ (ignore dashed edges) in Fig.~\ref{fig-motivation-example},
 \grouprec returns the team found in Example~\ref{exm-rsimulation} as its only answer.

\sstab(2) For $P_1$ and $G_1$ (with dashed edges), \grouprec returns the team in (1) as the best,
and the connected component containing \kw{PM_2} with its density = 1 as the second best.
\end{example}

\stitle{Correctness \& complexity analyses}. The correctness of algorithm \grouprec is assured by the following.
(1) There is a unique perfect subgraph in each ball of $G$.
(2) The correctness of \rgraphsim can be verified along the same lines as for simulation~\cite{infsimu95}.
(3) The list $L_{k}$ always maintains the current top-$k$ teams with largest density.
It takes $O(|V||P||G|)$ time to compute team simulation for all balls,
and $O(|V|\cdot(|V_{G_s}|^2+|G_s|))$ time to maintain the top-$k$ teams.
Thus \grouprec is in $O(|V|\cdot(|P||G|+|V_{G_s}|^2))$ time.
}

\subsection{Dynamic Top-k Team Formation}
\label{subsec-dynteamF}

We now introduce dynamic top-$k$ team formation.

\stitle{Pattern updates ($\Delta P$)}. There are five types of pattern updates:
(1) {\em edge insertions} connecting nodes in $P$,
(2) {\em edge deletions} disconnecting nodes in $P$,
(3) {\em node insertions} attaching new nodes to $P$,
(4) {\em node deletions} removing nodes from $P$, and
(5) {\em capacity changes} adjusting the node capacities in $P$,
while $P$  remains connected in all cases.

\stitle{Data updates ($\Delta G$)}. There are four types of data updates,
defined along the same lines as the first four types of pattern updates.
Further, different from pattern updates, there is no need to keep $G$ connected for data updates.

\stitle{Dynamic top-$k$ team formation}. Given pattern $P$, data graph $G$, positive integers $r$ and $k$, the list $L_{k}(P,G)$ of top-$k$ perfect subgraphs for $P$ in $G$,
a set of pattern updates $\Delta P$ and a set of data updates $\Delta G$,
the {\em dynamic top-k team formation} problem, denoted by \dynteamF{(P, G, k, L_{k}, \Delta P, \Delta G)},
is to find a list of $k$ perfect subgraphs with the top-$k$ largest density for $P\oplus\Delta P$ in $G\oplus\Delta G$, via team simulation.

Here $\oplus$ denotes applying changes $\Delta P$ to $P$ and $\Delta G$ to $G$, and
 $P\oplus\Delta P$ and $G\oplus\Delta G$ denote the updated pattern and data graphs.
 It is worth mentioning that \dynteamF{} covers a broad range of dynamic situations,
\ie handling continuously separate and simultaneous pattern and data updates.

\eat{
\begin{example}
\label{exm-dyn-team}
Recall $P_1$, $G_1$, $\Delta P_1$ and $\Delta G_1$ in Example \ref{exm-motivation-inc} (3), and set $r=2$ and $k=2$.
When get the top-$2$ teams for $P_1$ in $G_1$, \ie $L_k(P_1,G_1)$,
\dynteamF{} is to find the top-$2$ teams for $P_{1}\oplus\Delta P_{1}$ in $G_{1}\oplus\Delta G_{1}$ via team simulation.
\end{example}}

\eat{
\subsection{Optimization Techniques}
\label{subsec-batch-opt}
We next develop two optimizations for algorithm \grouprec.

\stitle{Density based filtering}. This technique allows algorithm \grouprec to compute the top-$k$ perfect subgraphs \wrt $P$ without necessarily computing team simulation for all balls in $G$. The important issue is how can tell whether a ball has the possibility to have one of the final top-$k$ results.
The idea is, given a ball $\ball{[v,r]}$ in $G$, we calculate the upper bound of $\density{\hat{G}_s}$,
where $\hat{G}_s$ is a subgraph of $\ball{[v,r]}$.
If the bound is larger than the current $k$-th result, \ie there is possibility the final answer resides in the ball, compute team simulation in it \wrt $P$;
Otherwise, skip the ball to the remaining balls to avoid redundant team simulation computing.

The problem here is how to efficiently compute the upper bound of $\density{\hat{G}_s}$ for each ball in $G$.
As the best densest subgraph algorithms are in $O(|\ball{[v,r]}|^3)$ time \cite{maximumDenseSubgraph}, which is costly,
we utilize an important result in~\cite{EVMK12}, shown as follows.

\begin{lemma}
\label{lemma-approximation-bound}
Let $\density{H_{c}}$ and $\density{H_{d}}$ be the density of the maximum core  $H_{C}$  and the  densest subgraph $H_{d}$ of graph $H$. Then (1) $\density{H_{c}}\leq \density{H_{d}} \leq 2*\density{H_{c}}$; and (2) there exists an algorithm that computes $\density{H_{c}}$ in $O(|E_H|)$ time~\cite{EVMK12}.
\end{lemma}

Here the maximum core $H_{C}$ of a graph $H$ is a subgraph of $H$ whose node degree is at least $\rho$, where $\rho$ is the maximum possible one. By Lemma~\ref{lemma-approximation-bound}, we use $2*\density{H_{c}}$ as the density upper bound for filtering unnecessary balls.

\stitle{Pattern minimization}.
Minimizing patterns is an effective way for improving the efficiency of querying graphs.


We say two pattern graphs $P$ and $P'$ are {\em equivalent via team simulation}, denoted by $P\equiv P'$, iff they return the same result on any data graph $G$ via team simulation. We say $P$ is {\em minimum} if for any other $P'$ such that $P\equiv P'$, $|P|\leq |P'|$.

\begin{theorem}
\label{thm-pattern-minimization}
For any pattern $P$, (1) there exists a unique minimum equivalent pattern $P_m$ via team simulation;
and (2) there exists an algorithm finding $P_m$ in $(|P|^2)$ time.
\end{theorem}

We propose an algorithm to minimize patterns (\ie\ \minp in the Appendix), which
 computes the maximum match relation $\Reps$ \wrt graph simulation by treating $P$ as both pattern and data graphs,
combines equivalent nodes \wrt $\Reps$, and merges capacity bounds on equivalent nodes.

\begin{example}
Takes as input the pattern $P_3$ shown in Fig.~\ref{fig-consistency-example}.
By Proposition~\ref{prop-pattern-consistency}, it determines that pattern $P_3$ is satisfiable.
By algorithm \minp, it constructs the minimum equivalent pattern graph $P_{3m}$ of $P_3$, shown in Fig.~\ref{fig-consistency-example}.
\end{example}

Note that algorithm \grouprec is finally incorporated with the above two optimization techniques.
}




\eat{
{\bf 1. more details are needed.}

{\bf 2. Add an example.}

{\bf 3. Add Correctness and Complexity.}

{\bf 4. why not for k: space issues for preprocessing}

\textbf{to add: (1) the analyses and experiments on the space of balls: (a) directly store all balls (nodes and the edges for border nodes), (b) using reference method~\cite{NavlakhaRS08} to compress the balls}

This is to justify the solution that stores no balls.

\textbf{to add: (2) the analyses and experiments on the counter filters:}

This is to justify the solution of bloom filters and bitmap index.
}

\section{Finding Top-k Teams}
\label{sec-tsimAlg}

\begin{figure}[tb!]
	\begin{center}
		\includegraphics[scale=1.38]{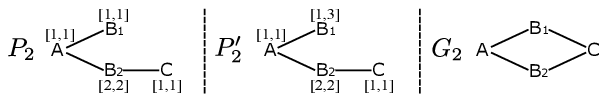}
	\end{center}
	\vspace{-3ex}
	\caption{Pattern satisfiability}
	\vspace{-4ex}
	\label{fig-consistency-example}
\end{figure}


In this section, we develop a batch algorithm for top-$k$ team formation.
We first study the pattern satisfiability problem for team simulation, then introduce two optimization techniques, and finally we present our batch algorithm.

\subsection{Pattern Satisfiability}
Different from graph simulation \cite{infsimu95} and its extensions \cite{FanLMTWW10,MaCFHW14},
there exist patterns that cannot match any data graph via team simulation, due to the presence of capacity constraints on patterns. We illustrate this with an example.

\begin{example}
\label{exm-consistency}
\ni(1) For pattern $P_2$ in Fig.~\ref{fig-consistency-example},
one can verify that there exist no data graphs $G$ such that $P_2 \eeps G$ because (a) for any nodes $v$ in $G$, if $v$ matches with the node labeled with $B_2$, then it must match with the node labeled with $B_1$, and, hence, (b) the capacity upper bound on $B_1$ should not be less than the lower bound on $B_2$.

\sstab(2) However, pattern $P'_2$ in Fig.~\ref{fig-consistency-example} is satisfiable as $P'_2 \eeps G_2$, and pattern $P_1$ in Fig.~\ref{exm-motivation} is also satisfiable as $P_1 \eeps G_1$.
\end{example}


We say that a pattern $P$ is {\em satisfiable} iff there exists a data graph $G$ such that $P$ matches $G$ via team simulation, \ie $P \eeps G$.
The good news is that checking the satisfiability of pattern graphs can be done in low polynomial time.

\begin{prop}
\label{prop-pattern-consistency}
The satisfiability of patterns $P$ can be checked in $O(|P|^2)$ time.
\end{prop}

By treating $P$ as both data and pattern graphs, compute the maximum match relation $M$ in $P$ for $P$, via graph simulation.
Then pattern $P$ is satisfiable iff for each $(u, v)\in M$ with the capacity bounds $[x_u, y_u]$ on $u$ and $[x_v, y_v]$ on $v$, respectively, $x_v \leq y_u$ holds. Observe that the size of $M$ is bounded by $|P|^2$, and pattern graphs are typically small.

By Proposition~\ref{prop-pattern-consistency}, we shall consider  satisfiable pattern graphs only  in the sequel.

\eat{
Pattern graphs are typically small, and we only consider  satisfiable patterns in the sequel, by Proposition~\ref{prop-pattern-consistency}.
}

\subsection{Batch Algorithm}

We then introduce two techniques for optimizing the computation of team simulation.

\stitle{Handling radius varied balls}. \teamF{} is to find top-$k$ teams within balls $\ball{[v, t]}$, where $v \in V$ and $t \in [1,r]$.
However, it is very costly to construct all $r|V|$ balls, and to compute perfect subgraphs in all of them.
Indeed it is also not necessary, and it only needs to construct and compute the matches for a number of $|V|$ balls, \ie the set of balls $\ball{[v, r]}$
where $v \in V$ and radius is $r$, and then incrementally computes the perfect subgraphs for balls $\ball{[v, t]}$ ($t \in [1,r-1]$) from the match graphs for ball $\ball{[v, r]}$, as shown below.

\vspace{-1ex}
\begin{theorem}
	\label{thm-tsim-radius}
	Given $P$, ball $\ball{[v, r]}$ and $\ball{[v, t]}$ ($t \in[1,$ $r-1]$) in $G$,
	(1) if $P\eps \ball{[v, t]}$, then $P\eps \ball{[v, r]}$; and
	(2) if $G_s$ (resp. $G'_s$) is the match graph \wrt the maximum match relation $M$ (resp. $M'$) in $\ball{[v, r]}$ (resp. $\ball{[v, t]}$) for $P$ via graph simulation, then $M' \subset M$, and $G'_s$ is a subgraph of $G_s$.	
\end{theorem}
\looseness=-1

When we have the match graph $G_s$ in $\ball{[v, r]}$ for $P$ via graph simulation, to compute the perfect subgraph in $\ball{[v, t]}$ ($t \in [1,r-1]$) for $P$ via team simulation, we need to
(1) first identify the subgraph $G_s^{t}$ in $G_s$ belonging to $\ball{[v, t]}$, which can be easily identified in the process for constructing $\ball{[v, r]}$ without extra computation;
(2) check whether $G_s^{t}$ is already a match graph for $P$ in $\ball{[v, t]}$ via graph simulation; if not, remove the unmatched nodes and edges from $G_s^{t}$ until find the match graph $G'_s$ for $P$ in $\ball{[v, t]}$. This can be achieved by executing an efficient incremental process in~\cite{FanWW13-tods}; and
(3) finally check whether capacity bounds are satisfied. If so, $G'_s$ is the perfect subgraph in $\ball{[v, r]}$ for $P$ via team simulation.

\stitle{Density based ball filtering}. We further reduce the number of balls to speedup the process by adopting the density based filtering technique.
The key idea is to tell whether a ball is possible to produce one of the final top-$k$ matches.

Given a ball $\ball{[v,r]}$, we compute the density upper bound  $\density{\hat{G}_s}$,
where $\hat{G}_s$ is a subgraph of $\ball{[v,r]}$.
If the bound is larger than the density of the current $k$-th result, \ie there is a possibility for the ball;
Otherwise, the ball is simply ignored to avoid redundant computations.

The trick part is how to efficiently compute the upper bound of $\density{\hat{G}_s}$ for each ball in $G$.
As the best densest subgraph algorithms are in $O(|\ball{[v,r]}|^3)$ time \cite{maximumDenseSubgraph}, which is costly,
we utilize an important result in~\cite{EVMK12}, shown below.

\begin{lemma}
	\label{lemma-approximation-bound}
	Let $\density{H_{c}}$ and $\density{H_{d}}$ be the density of the maximum core  $H_{C}$  and the  densest subgraph $H_{d}$ of graph $H$. Then (1) $\density{H_{c}}\leq \density{H_{d}} \leq 2*\density{H_{c}}$; and (2) there exists an algorithm that computes $\density{H_{c}}$ in $O(|E_H|)$ time~\cite{EVMK12}.
\end{lemma}

Here the maximum core $H_{C}$ of a graph $H$ is a subgraph of $H$ whose node degree is at least $\rho$, where $\rho$ is the maximum possible one. By Lemma~\ref{lemma-approximation-bound}, we use $2*\density{H_{c}}$ as the density upper bound for filtering unnecessary balls.

We are now ready to present our batch algorithm for \teamF{}.

\stitle{Algorithm \grouprec.} As shown in Fig.~\ref{alg-grouprec}, it takes input as $P$, $G$, and two integers $r$ and $k$, and outputs the top-$k$ densest perfect subgraphs in $G$ for $P$.
It firstly checks whether $P$ is satisfiable (line 1).
If so, for each ball $\ball{[v,r]}$ in $G$, it computes the maximum core $\ball{}_{C}$ of $\ball{[v,r]}$, and checks whether the
density based ball filtering condition holds (lines 3-6).
If so, it skips the current ball, and moves to the next one;
otherwise, it computes the perfect subgraph $G_{s}$ of $P$ in $\ball{[v,r]}$ via team simulation by invoking \rgraphsim (line 7, see full version~\cite{fullvldb18}),
an adaption from graph simulation~\cite{infsimu95,FanLMTWW10} and checking capacity bounds (line 8).
It then computes perfect subgraphs $G'_{s}$ of $P$ in inner balls $\ball{[v,t]}$ by invoking \incsim, an extension of the data incremental algorithms in \cite{FanWW13-tods} and checking capacity bounds (lines 9-11).

\stitle{Correctness \& complexity analyses}. The correctness of \grouprec is assured by the following.

\sstab(1) The correctness of \rgraphsim (resp. \incsim) can be verified along the same lines as graph simulation~\cite{infsimu95} (resp. incremental simulation \cite{FanWW13-tods}).

\sstab(2) Theorem~\ref{thm-tsim-radius} and Lemma~\ref{lemma-approximation-bound}.
It takes $O(|P|^2)$ to check pattern satisfiability, $O(|V||P||G|)$ to compute team simulation, $O(r|V||V_{P}||E|)$ to incrementally compute matches in inner balls, and $O(|V||E|)$  to compute the density of the maximum core for $|V|$ balls.
Thus \grouprec is in $O(|P|^2+|V||P||G|+r|V||V_{P}||E|)$. However, actual time is much less due to density based ball filtering and that  $O(r|V||V_{P}||E|)$ is the worst case complexity for incremental process, while $r$ is small, \ie 2 or 3.

\begin{figure}[t!]
	\begin{center}
		{\small
			\myhrule
			\vspace{-3ex}
			\mat{0ex}{
				\sstab {\sl Input:\/} $G(V$, $E)$, $P(V_P$, $E_P)$, and positive integers $r$ and $k$.\\
				{\sl Output:\/} Top-$k$ densest teams.\\
				\sstab \bcc \hspace{1ex}\= \If $P$ is unsatisfiable \Then \Return $nil$;\\
				\icc \> $L_{k} := \emptyset$;\\
				\icc \> \For each ball $\ball{[v,r]}$ in $G$ \Do\\
				\icc \>\hspace{2ex}\= compute the maximum core $\ball{}_{C}$ of the ball $\ball{[v,r]}$;\\
				\icc \>\> \If 2*$\density{\ball{}_C} \le$  the density of the $k$-th result in $L_{k}$ \Then\\
				\icc  \>\>\hspace{2ex}\= \Continue;\\
				\icc \>\> $G_s$ := \rgraphsim$(P, \ball{[v,r]})$; \\
				\icc \>\> If $G_s$ satisfies capacity bounds on $P$ \Then Insert $G_s$ into $L_{k}$;\\
				\icc \>\> \For each ball $\ball{[v,t]}$ with $t\in[1,r-1]$ \Do\\
				\icc \>\>\hspace{2ex}\= $G'_s$ := \kw{incSim}$(G_s, P, \ball{[v,t]})$;\\
				\icc \>\>\> If $G'_s$ satisfies capacity bounds on $P$ \Then Insert $G'_s$ into $L_{k}$;\\
				\icc \>\, \Return $L_{k}[0:k-1]$.
			}
			
			\vspace{-4ex} \myhrule
		}
	\end{center}
	\vspace{-3ex}
	\caption{Algorithm \grouprec}
	\label{alg-grouprec}
	\vspace{-4ex}
\end{figure}

\newcommand{\changedinc}{\kw{Changed}}
\newcommand{\cut}{\kw{CUT}}
\newcommand{\imr}{\kw{IMR}}
\newcommand{\imrs}{\kw{IMRs}}
\newcommand{\fb}{\kw{FBM}}
\newcommand{\bs}{\kw{BS}}
\newcommand{\fs}{\kw{FS}}
\newcommand{\bfc}{\kw{BF}}
\newcommand{\upl}{\kw{UP}}
\newcommand{\fbmatstruct}{\kw{FBMatStruct}}
\newcommand{\matchindex}{\kw{MatStaIndex}}
\newcommand{\matchimr}{\kw{MatchIMR}}
\newcommand{\for}{\kw{FOR}}
\newcommand{\while}{\kw{WHIlE}}
\newcommand{\allmatch}{\kw{mat}}
\newcommand{\affnode}{\kw{affnode}}

\newcommand{\ms}{\kw{SHOUDfs}}
\newcommand{\ballfilter}{\kw{SHOUDbfc}}

\newcommand{\dens}{\kw{den}}
\newcommand{\identifyaffball}{\kw{IdABall}}
\newcommand{\patedgeinsert}{\kw{patEIns}}
\newcommand{\incmatch}{\kw{IncMatch}}
\newcommand{\wmatchindex}{\kw{wMatStaCode}}
\newcommand{\cflag}{\kw{cflag}}
\newcommand{\dflag}{\kw{dflag}}
\newcommand{\gflag}{\kw{gflag}}
\newcommand{\rflag}{\kw{rflag}}

\newcommand{\incgrpat}{\kw{kPatIncGPM}}
\newcommand{\incgrdata}{\kw{kDataIncGPM}}
\newcommand{\dyngr}{\kw{kDynGPM}}
\newcommand{\affballx}{\kw{AffB}}
\newcommand{\affballsx}{\kw{AffBs}}
\newcommand{\affballacc}{\kw{AffBall^{acc}}}
\newcommand{\affballaccs}{\kw{AffBalls^{acc}}}
\newcommand{\affballimr}{\kw{AffBall^{imr}}}
\newcommand{\affballimrs}{\kw{AffBalls^{imr}}}
\newcommand{\patinc}{\kw{dynamicPG}}
\newcommand{\datainc}{\kw{dynamicDG}}
\newcommand{\optpatinc}{\kw{SHOULDoptinc}}
\newcommand{\inc}{\kw{dynamic}}
\newcommand{\matchs}{\kw{R}}
\newcommand{\comb}{\kw{combine}}
\newcommand{\optinc}{\kw{optDynamic}}

\newcommand{\incp}{\kw{dynamicP}}
\newcommand{\incd}{\kw{dynamicG}}

\section{A Unified Incremental Solution}
\label{sec-dynamictopk}

In this section, we first analyze the challenges and design principles of dynamic top-$k$ team formation,
and then develop a unified incremental framework for \dynteamF.
For convenience, the notations used are summarized in Table~\ref{tab-notation}.

\subsection{Analyses of Dynamic Team Formation}

By Theorem~\ref{thm-tsim-radius}, pattern $P$ matches a ball $\ball{[v, t]}$ ($t \in [1,r-1]$), only if $P$ matches ball $\ball{[v, r]}$ via graph simulation,
and the match results for $\ball{[v, t]}$ can be derived from the matches for $\ball{[v, r]}$.
Therefore, the key of the incremental computation is to deal with the balls $\ball{[v, r]}$ with radius $r$.
In the sequel, a ball has a radius $r$ by default.


\eat{
Like data incremental problems, it would be nice to have {\em bounded incremental algorithms}~\cite{Reps96} for pattern incremental problems, which are a class of incremental algorithms whose time complexity depends on the sizes of $P$ and $\changedinc$, which is the changes to the input and output only (\ie $\Delta P$ and the difference between $PG_k(P, G)$ and $PG_k(P\oplus\Delta P, G)$),
and is independent of the size of $G$ and $PG_k(P, G)$.
However, due to the inherent difference between pattern increments and data increments, and the presence of top-$k$ semantic, they are not available for \incgrpat, as shown below.
We say an incremental problem is bounded if it has a bounded incremental algorithm described as above, and is unbounded otherwise.

Similar to the negative results in \cite{FanLMTWW10,FanWW13-tods}

The following results show that even we consider pattern or data increments alone, traditional bounded incremental algorithms are not available for \dyngr already.

%
%

}

We first analyze the inherent computational complexity of the dynamic top-$k$ team formation.

\stitle{Incremental complexity analysis}. As observed in~\cite{Reps96,inc-survey}, the complexity of incremental algorithms should be measured by the size $|\aff|$ of {\em the changes} in the input and output, rather than the entire input, to measure the amount of work essentially to be performed for the problem.

An incremental problem is said to be {\em bounded} if it can be solved by an algorithm whose complexity is a function of $|\aff|$ alone, and is {\em unbounded}, otherwise. Unsurprisingly, the dynamic top-$k$ team formation problem is unbounded, similar to the other extensions of graph simulation \cite{FanLMTWW10,FanWW13-tods}.

\begin{prop}
\label{thm-inc-grouprec-pat}
The \dynteamF{} problem is unbounded, even for $k$ = 1 and unit pattern or data updates.
\end{prop}

We then illustrate the impact of pattern and data updates on the matching results with an example.

\begin{example}
\label{exm-pattern-challenge}
Continue Example~\ref{exm-motivation-inc} with $\Delta P_1$ and $\Delta G_1$.


\sstab
(1) For $\Delta P_1$, $\ball{[\kw{PM_1}, 2]}$ already matches $P$, and may produce more matched nodes for $P_1\oplus \Delta P_1$,
thus a re-computation for perfect subgraphs is needed.
For all other balls, $\Delta P_1$ may turn unmatched nodes to matched and may produce perfect subgraphs,
thus re-computation is also needed.

\sstab
(2) For $\Delta G_1$, it produces a new perfect subgraph for $P$ in $G_1\oplus\Delta G_1$,
\ie the connected component having \kw{PM_2}.
\end{example}


\begin{table}[tb!]
	\begin{center}
		\begin{small}
			\scriptsize
			\begin{tabular}{|c|l|}
				\hline
				{\bf Notations}             &  {\bf Description}     \\
				\hline\hline
				$P,G$                     &  pattern and data graphs       \\ \hline
				$\ball{[v, r]}$              &  a ball in $G$ with center node $v$ and radius $r$  \\ \hline
				$L_k(P, G)$                  &  the list of top-$k$ perfect subgraphs in $G$ for $P$  \\ \hline
				$\Delta P, \Delta G$       &  pattern and data updates       \\ \hline
				$\oplus$                    &  applying updates $\Delta P$ and $\Delta G$ to $P$ and $G$       \\ \hline
				${\cal P}_{h}=\{P_{fi}, C\}$     &  pattern fragmentation: $h$ fragments and cut    \\ \hline
				$\affballsx$                &  affected balls       \\ \hline
				$M(P_{fi}, \ball{[v,r]})$    &  the maximum match relation in $\ball{[v,r]}$ for $P_{fi}$   \\ \hline
				$\tilde{M}(P,G)$            &  fragment-ball matches (auxiliary structure)     \\ \hline
				$\fs$, $\bs$              &  fragment status, ball status (auxiliary structure) \\    \hline
				$\fb$                       &  fragment-ball-match index, containing $\fs, \bs$ \\    \hline
				$\bfc$, $\upl$              &  ball filter, update planner (auxiliary structure) \\    \hline
			\end{tabular}
			\vspace{-1ex}
		\end{small}
		\caption{Notations}
		\label{tab-notation}
		\vspace{-6ex}
	\end{center}
\end{table}

We finally discuss the challenges and principles of designing incremental algorithms for \dynteamF{} from three aspects.


\stitle{(1) Impacts of pattern and data updates}.
Beyond Proposition~\ref{thm-inc-grouprec-pat} and Example~\ref{exm-pattern-challenge}, one can also verify that (a) unit pattern updates are likely to result in the entire change in previous results, such that all balls need to be accessed and all matches need to be re-computed,
and (b) the impact of data updates can also be global, such that the entire data graph may need to be accessed to re-compute matches.
Hence, the key is to identify and localize the impacts of pattern and data updates.


\stitle{(2) Maintenance of auxiliary information}.
Auxiliary data on intermediate or final results for $P$ in $G$ are typically maintained for incremental computation \cite{Reps96,FanWW13-tods}.
How to design light-weight and effective auxiliary structures is critical.
One may want to store $M(P,G)$, the match relations of $P$ for all balls in $G$,
as adopted by existing incremental pattern matching algorithms for data updates \cite{FanWW13-tods}.
However, the impact of $\Delta P$ is global, as shown in Example~\ref{exm-pattern-challenge}.
By storing $M(P,G)$, for pattern edge/node deletions, it has to recompute matches for all balls, \ie the entire $M(P,G)$.
Thus, storing $M(P,G)$ could be useless, not to mention $L_{k}(P,G)$, the list of top-$k$ perfect subgraphs for $P$ in $G$ \wrt $M(P,G)$.

\stitle{(3) Support of continuous pattern and data updates}.
A practical solution should support continuous pattern and data updates, separately and simultaneously, which further increases difficulties on the design of auxiliary data structures and incremental algorithms.

\subsection{A Unified Incremental Framework}
\label{subsec-framework}

Nevertheless, we develop an incremental approach to handling pattern and data updates in a unified framework, by utilizing {\em pattern fragmentation} and {\em affected balls} to localize the impacts of pattern and data updates, and to reduce the cost of maintaining auxiliary structures and computations.


\stitle{(I) Localization with pattern fragmentation}.
We say that \{$P_{f1}(V_{f1}, E_{f1})$, $\ldots$, $P_{fh}(V_{fh}, E_{fh})$, $C$\} is an {\em $h$-fragmentation} of pattern $P(V_{P}$, $E_{P})$, denoted as ${\cal P}_{h}$,
if (1) $\bigcup_{i=1}^{h}V_{fi}=V_{P}$, (2) $V_{fi}\cap V_{fj} = \emptyset$ for any $i\ne j\in[1, h]$,
(3) $E_{fi}$ is exactly the edges in $P$ on $V_{fi}$, and (4) $C=E_P \setminus (E_{f1}\cup\ldots\cup E_{fh})$.

We also say $P_{fi}$ ($i\in[1,h]$) as a {\em fragment} of $P$, and $C$ as a {\em cut} of $P$, respectively.

Observe that by pattern fragmentation, a pattern update on $P$ is either on a fragment $P_{fi}$ or on the cut $C$ of $P$, and, in this way, the impact of pattern updates is localized. Moreover, graph simulation holds a nice property on pattern fragmentation, as shown below.

\vspace{-0.5ex}
\begin{theorem}
\label{thm-compose}
Let $\{P_{f1},\ldots,P_{fh}\}$ be an $h$-fragmentation of pattern $P$.
For any ball $\ball{}$ in $G$, let $M_i$ ($i\in[1,h]$) be the maximum match relation in $\ball{}$ for $P_{fi}$ via graph simulation,
and $M$ be the maximum match relation in $\ball{}$ for $P$ via graph simulation, respectively,
then $M\subseteq\bigcup_{i=1}^{h}M_{i}$.
\end{theorem}

We also say that $M_i$ is a {\em partial match relation} in ball $\ball{}$ for $P$ via graph simulation.
By the nature of graph simulation~\cite{infsimu95}, $\bigcup_{i=1}^{h}M_{i}$ is actually an intermediate result of $M$.
Once we have the maximum match relation $M$ for $P$ in $\ball{}$, via graph simulation, we can further produce the result for $P$ in  $\ball{}$ via team simulation, by a capacity check.

That is, based on pattern fragmentation, we maintain
an auxiliary structure for storing fragment-ball matches for incremental computations,
\ie $\tilde{M}(P,G)$ \wrt ${\cal P}_{h}$ that is the maximum match relations for all pattern fragments of $P$ in all balls of $G$, via graph simulation.
Moreover, its space cost is light-weight, as will be shown in the experimental study.


By storing $\tilde{M}(P,G)$, we have $\bigcup_{i=1}^{h}M_{i}$ for each ball $\ball{}$,
and we can simply update $M_i$ while leaving other parts untouched. That is, we indeed compute for $P_{fi}\oplus\Delta P(\ball{})$, instead of $P\oplus\Delta P(\ball{})$, and combine all $P_{fi}\oplus\Delta P(\ball{})$ to derive $P\oplus\Delta P(\ball{})$.
Even better, the updates $\Delta P$ on the cut $C$ of $P$ only involve with a simple combination process, avoiding the computation for any pattern fragments.

For a better incremental process, we typically want (1) to avoid skewed updates by balancing the sizes of all fragments, and (2) to minimize the efforts to assemble the partial matches of all fragments.
Thus we define and investigate the {\em pattern fragmentation} problem.

Given pattern $P$ and a positive integer $h$, it is to find
an $h$-fragmentation of $P$ such that both $\max(|P_{fi}|)$ ($i\in[1,h]$) and $|C|$ are minimized.
Intuitively, the bi-criteria optimization problem partitions a pattern into $h$ components of roughly equal size while minimizing the cut size.

\eat{
Given $P$ and a positive integer $h$, the {\em pattern fragmentation}  problem is to find
an $h$-fragmentation of $P$ such that both $\max(|P_{fi}|)$ ($i\in[1,h]$) and $|C|$ are minimized.
Intuitively, we want (1)  to avoid skewed updates by making all fragments roughly the same size and (2) to  minimize the efforts to assemble the partial matches of all fragments.
}

The problem is intractable, as shown below.

\vspace{-0.5ex}
\begin{prop}
	\label{prop-fragmentation}
	The pattern fragmentation problem is \NP-complete, even for $h$ = 2.
\end{prop}

However, $P$ and $h$ are typically small in practice~\cite{FanLMTWW10}, \eg $|P|=15$ and $h=3$.
In light of this, we give a heuristic algorithm, denoted by \kw{PFrag}, for the problem, and is shown in the
full version~\cite{fullvldb18}.
\kw{PFrag} works by connecting pattern fragmentation to the widely studied {\sc $(k, \nu)$-Balanced Partition} problem \cite{AndreevR06},
which is not approximable in general, but has efficient and sophisticated heuristic algorithms~\cite{metis-KarypisK98a}.

\stitle{(II) Localization with affected balls (\affballsx)}.
We further localize the impact of pattern and data updates  with {\em affected balls} to avoid unnecessary computations.

We say that a ball in $G$ is {\em affected} \wrt an incremental algorithm \algorithm{A}, and pattern and data updates,
if \algorithm{A} accesses the ball again.
We use $||\affballsx||$ and $|\affballsx|$ to denote the cardinality and total size of \affballsx, respectively.


%
Indeed, \affballsx are those balls with a possibility to have final results \wrt $\Delta P$ and $\Delta G$.
We only access \affballsx, and ignore the rest balls.
Specifically, (1) for $\Delta P$, it allows us to avoid computing updated partial relations for an updated fragment in every ball;
and (2) for $\Delta G$, the locality property of team simulation supports to localize the update impacts to a set of balls whose structures are changed by $\Delta G$.
\looseness=-1


\stitle{(III) Algorithm framework}. We now provide a unified incremental algorithm to handle both pattern and data updates,
based on pattern fragmentation and affected balls.

Given pattern $P$ with its $h$-fragmentation ${\cal P}_{h}$, data graph $G$, two integers $r$ and $k$,
and auxiliary structures (to be introduced in Section \ref{sec-IncAlg}) such as the partial match relations for all pattern fragments and all balls (radius $r$),
algorithm \inc  consists of three steps for $\Delta P$ and $\Delta G$, as follows.

\etitle{(1) Identifying \affballsx}.
Algorithm \inc invokes two different procedures to identify \affballsx for separate $\Delta P$ or $\Delta G$, respectively.
For  simultaneous $\Delta P$ and $\Delta G$, \inc takes the union of the \affballsx produced by  the two procedures.

\etitle{(2) Update partial match relations in \affballsx}.
For a ball affected by $\Delta P$, \inc updates the partial match relations for the updated pattern fragments with incremental computation;
For a ball affected by $\Delta G$, \inc updates the partial match relations for all pattern fragments;
And, for a ball affected by both $\Delta P$ and $\Delta G$, \inc follows the same way as it does for $\Delta G$ only.
Meanwhile, auxiliary structure \fb (to be seen shortly) is updated for handling continuously separate and simultaneous pattern and data updates.

\etitle{(3) Combining partial match relations}. \inc combines all partial relations for a subset of \affballsx and computes the top-$k$ perfect subgraphs within them and their inner balls.

Observe that \inc handles pattern and data updates, separately and simultaneously, in a unified way.

\eat{
Theorem~\ref{thm-inc-grouprec-pat} and ~\ref{thm-inc-grouprec-data} show that, conventional bounded incremental algorithms are not available for $\incgrpat$ and $\incgrdata$ even for unit update only.
The main reasons are three challenges introduced by pattern/data incremental and the top-$k$ semantic, summarized as follows.

\sstab {\bf Challenge (1): impacts of pattern or data changes}.
The first challenge is how to identify and restrict, if possible, the impacts of pattern or data changes. This is nontrivial.
(a) Due to the nature of pattern changes, a unit update on pattern graph may have impact on all perfect graphs in $PG(P,G)$, or even add new matches. Indeed, for any algorithm takes input as $P$, $G$, $k'$, $PG_k(P,G)$ and $\Delta P$, it has to access entire $G$ (every ball in $G$), even when $\Delta P$ is a unit update, $k' = 1$ and $k = |V_G|$. That is, the impacts of a pattern update can be global.
(b) Data changes may also cause complicated impacts on both the ball structures and the number of balls in $G$. For one thing, the number of balls in $G$ may increase or decrease with new data changes; for another, the match relations in existing balls may have to be re-computed since the structure of balls may change.
These impacts may vary over updates, and require us to identify them.

\sstab {\bf Challenge (2): necessary auxiliary information}.
The second challenge comes from the uninformative in the output top-$k$ perfect graphs of the team simulation.
(a) Due to the top-$k$ semantics, $PG_k(P,G)$ contains only top-$k$ perfect graphs \wrt match relations in $M(P,G)$, which could be extremely uninformative even when $k$ is large, as any changes in the results may requires the knowledge of all perfect graphs to enforce the top-$k$ semantics.
(b) Worse still, for pattern changes, even all perfect subgraphs $PG(P,G)$ are precomputed and cached, it can provide no help for pattern incremental when there are pattern edge/node deletes, as indicated in example~\ref{exm-pattern-challenge}.
(c) Traditional incremental algorithms typically maintain some auxiliary data structures to accelerate the incremental process. However, the physical storage is limited in practice, and essential structures needed for team simulation may not be able to fit in the storage budget.
Indeed, we examined basic structures for team simulation and using real-life data to validate whether they are suitable as auxiliary structures.
(i) The first candidate is the balls, since balls of the data graph are the basic computation unit for team simulation. We test the cost of cache them. The results are discouraging.
It will take about 361GB, 437GB and 222GB for Citation, YouTube and Synthetic data ($10^7$ nodes) when $r$ is set to 3, and 338GB, 420GB and 222GB even when compression technique is adopted. This shows that precomputing and caching all balls in $G$ as auxiliary information is impractical.
(ii) Another choice is to precompute and store $M(P,G)$, as adopted by some pattern incremental algorithms~\cite{FanWW13-tods}.
However, as indicated in Challenge (1), the impact of $\Delta P$ is global. By storing $M(P,G)$, for edge/node inserts or capacity changes, the incremental process can get match results by updating all matches in $M(P,G)$ for all balls in $G$; worse still, for edge/node deletes, it has to recompute all matches in $M(P,G)$ for all balls in $G$. Therefore, storing $M(P,G)$ cannot improve the incremental process in many cases, worst still, it even takes extra cost to maintain $M(P,G)$.
}

\begin{figure*}[tb!]
	\begin{center}
		\subfigure{\label{fig-auxstr-index}
			\includegraphics[scale=0.585]{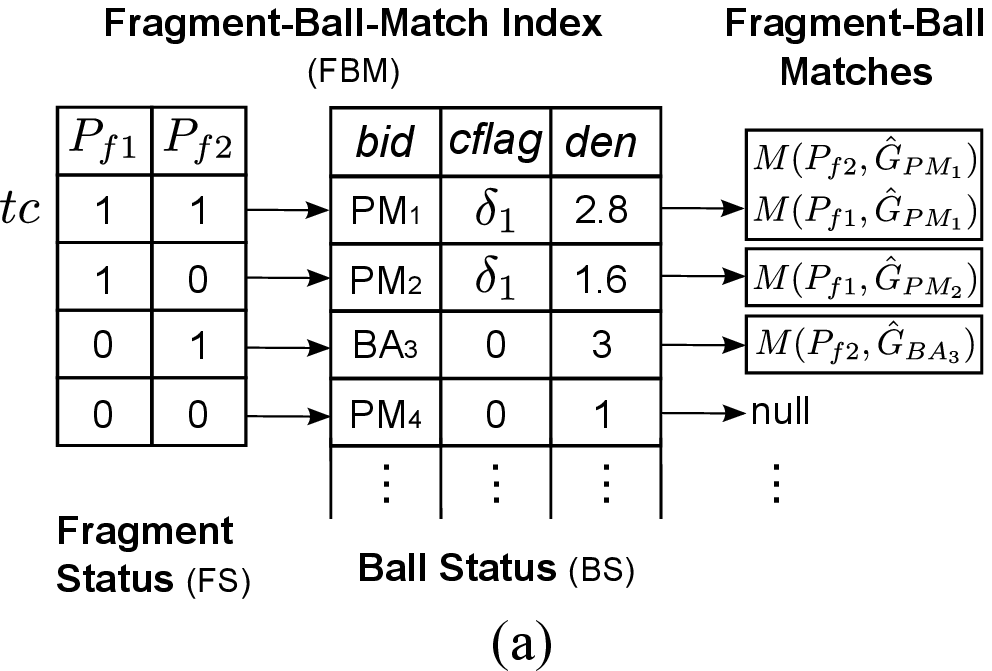}}
		\subfigure{\label{fig-inc-maintain-unit}
			\quad\quad \vrule \quad\quad
			\includegraphics[scale=0.585]{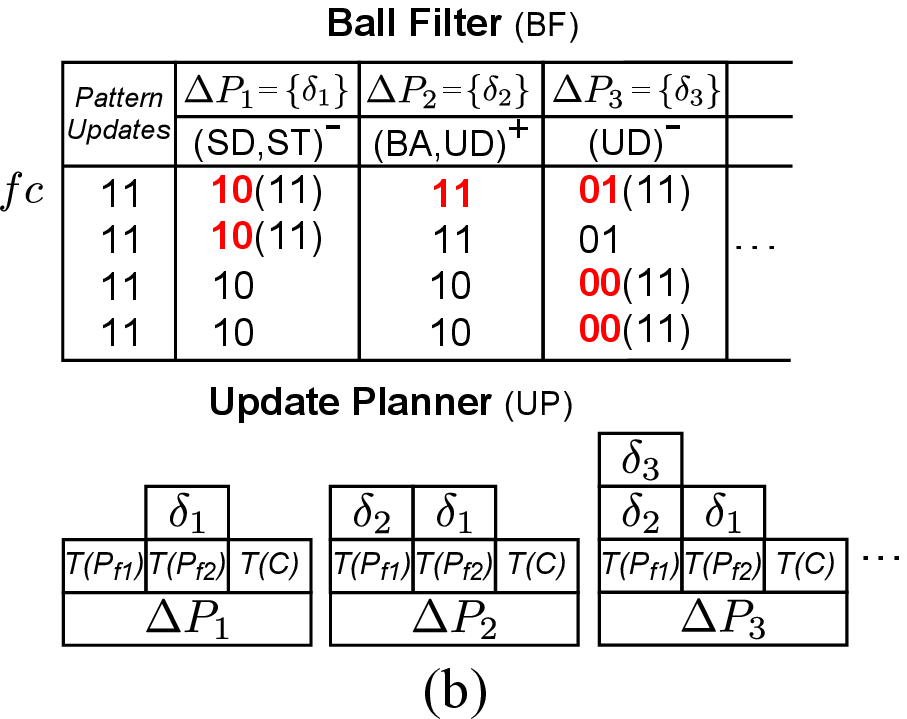}}
		\subfigure{\label{fig-inc-maintain-batch}
			\vrule \quad\quad
			\includegraphics[scale=0.585]{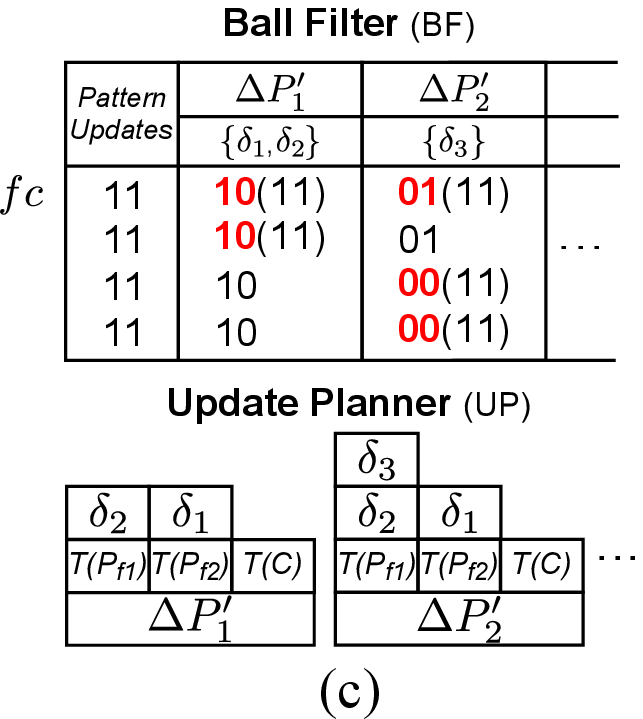}}
		\vspace{-2ex}
		\caption{Example auxiliary data structures}
		\label{fig-auxiliary-structures}
		\vspace{-5ex}
	\end{center}
\end{figure*}

\section{Incremental Algorithms}
\label{sec-IncAlg}

In this section, we introduce the details of our incremental algorithm \inc, including (a) auxiliary data structures,
(b) algorithms \incp and \incd to handle pattern and data updates, respectively, and (c) \inc by integrating \incp and \incd together.

\subsection{Auxiliary Data Structures}
\label{subsec-auxstr}

Auxiliary structures fall into two classes:
maintain partial matches and handle pattern incremental computing. Consider an $h$-fragmentation ${\cal P}_h$ = \{$P_{f1}, \ldots, P_{fh}, C$\} of pattern $P(V_P, E_P)$, data graph $G(V, E)$, and pattern updates $\Delta P$.

\stab(I) Data structures in the first class are as follows.

\etitle{(1) Fragment status (\fs)} consists of $2^h$ boolean vectors $(b_1,\ldots,b_h)$, referred to as {\em type code (tc)}, where $b_i$ $(i\in [1,h])$ is either 0 or 1. Recall that $h$ is very small, \eg 3.

We use \fs to classify the match status of balls in $G$ into $2^h$ types $\wrt$ ${\cal P}_h$.
For a ball with type code $(b_1,\ldots,b_h)$, $b_i$ is 1 iff $P_{fi}$ matches the ball via graph simulation.

\vspace{0.4ex}
\etitle{(2) Ball status (\bs)} consists of $|V|$ triples $(bid, cflag, den)$, such that $bid$ is the $id$ of a ball,
$cflag$ is the id of the latest processed unit pattern update for the ball (initially set to $0$),
and $den$ is the density upper bound of subgraphs in the ball.
\looseness=-1

We use \bs to store the basic information for balls in $G$.

\vspace{0.4ex}
\etitle{(3) Fragment-ball matches} of $P$ in $G$, denote as $\tilde{M}(P,G)$,
are $\bigcup_{i\in[1,h], v\in V} M(P_{fi},\ball{[v,r]})$,
such that $M(P_{fi},\ball{[v,r]})$ is the maximum match relation for $P_{fi}$ in ball $\ball{[v,r]}$, via graph simulation,
and there are in total $|V|$ balls.

Here $\tilde{M}(P,G)$ is used to store match relations for the pattern fragments of $P$ in all balls of $G$. Instead of storing a single $\tilde{M}(P, G)$, we organize $\tilde{M}(P, G)$ in terms of the match status between pattern fragments and balls, \ie\ \fs and \bs.

\vspace{0.4ex}
\etitle{(4) Fragment-ball-match index (\fb)} links \fs and \bs together, to form the {\em fragment-ball-match index}. Then \fb is linked to $\tilde{M}(P,G)$. The details are shown below.

For each record of ball $\ball{[v,r]}$ in \bs, (a) there is a link from its type code in \fs pointing to the record; and
(b) there is another link from the record to a set of $M({P}_{fi}, \ball{[v,r]})$ $(i\in [1,h])$ in $\tilde{M}(P,G)$,
if the type code with which the ball is associated has $b_i = 1$, \ie $M({P}_{fi}, \ball{[v,r]})$ is not empty.

Intuitively, \fb indexes the partial match relations $\tilde{M}(P,G)$ based on the match status of balls \wrt ${\cal P}_h$.

\begin{example}
\label{exa-matchindex}
Consider $P_1$ and $G_1$ (both without dashed edges) in Fig.~\ref{exm-motivation}, $r=2$, $k=2$, $h=2$,
auxiliary structures $\tilde{M}(P,G)$ and \fb\,that are shown in Fig.~\ref{fig-auxiliary-structures}(a).

\ni (1) Pattern $P_1$ is divided into fragments $P_{f1}$ and $P_{f2}$ by algorithm \kw{PFrag},
so there are $2^2=4$ type codes in \fs.

\ni (2) For balls linked with $tc$ $(1, 1)$, \eg ball $\ball{[\kw{PM_1},2]}$,
there are matches to both $P_{f1}$ and $P_{f2}$ in the ball.
Besides, there exist balls $\ball{[\kw{PM_2},2]}$, $\ball{[\kw{BA_3},2]}$ and $\ball{[\kw{PM_4},2]}$
linked with $tc$ $(1, 0)$, $(0, 1)$ and $(0, 0)$ respectively.
For simplicity, we use these 4 balls only in the following analysis.
\end{example}


These structures enforce a nice property as follows.

\begin{theorem}
\label{thm-framework-inc}
With $\tilde{M}(P,G)$ and \fb \wrt an $h$-fragmentation of $P$, given $\Delta P$ and $\Delta G$,
the incremental algorithm \inc processes $\Delta P$ and $\Delta G$ in time
determined by $P$, $\tilde{M}(P,G)$ and \affballsx, not directly depending on $G$.
\end{theorem}

We shall prove Theorem~\ref{thm-framework-inc} by providing specific techniques for \inc and analyzing its time complexity.

\stab(II) Data structures in the second class are as follows.

\vspace{0.5ex}
\etitle{(1) Ball filter  (\bfc)} consists of $2^{h}$ boolean vectors $(b_1$, $\ldots$, $b_h)$, referred to as {\em filtering code (fc)},
such that each $fc$ in \bfc corresponds to a type code $tc$ in \fs.
Each $b_i$ $(i\in [1, h])$ in an $fc$ of \bfc is initially set to $1$,
and is updated for each unit pattern update $\delta$ in $\Delta P$:
(a) when $\delta$ is an edge deletion or a node deletion to ${P}_{fi}$,
the $i$-th bit of all the $2^{h}$ filtering codes in \bfc is set to $0$; Otherwise, (b) \bfc remains intact.

\vspace{0.5ex}
\etitle{(2) Update planner (\upl)} consists of $h+1$ stacks $T({P}_{f1})$, $\ldots$, $T({P}_{fh})$, $T(C)$.
Stack $T(P_{fi})$ $(i\in [1,h])$ (resp. $T(C)$) records all unit updates in all arrived pattern updates $\Delta P_1$, $\ldots$, $\Delta P_{N}$ that are applied to fragment $P_{fi}$ (resp. $C$) of $P$.
Initially, all of them are empty, and are dynamically updated for each unit update in each coming set of  pattern updates.



\subsection{Dealing with Pattern Updates}
\label{subsec-Qinc}

We present algorithm \incp to handle pattern updates $\Delta P$, following the steps in Section~\ref{subsec-framework}, and
an early return optimization technique for \incp.



\stab{\bf (I) Identifying affected balls}.
We first develop procedure \identifyaffball to identify \affballsx with structures \fb and \bfc.

\stitle{Procedure \identifyaffball}.
Given an $h$-fragmentation ${\cal P}_{h}$ of $P$, $\Delta P$,
(1) it updates \bfc by processing all unit updates in $\Delta P$.
(2) For each $j\in [1, 2^h]$, it then executes a bitwise \kw{AND} operation (\&) between type code $tc_j$ of \fs in \fb and updated filtering code $fc_{j\Delta}$ in \bfc, \ie $tc_j\& fc_{j\Delta}$.
(3) Finally, if $tc_j\& fc_{j\Delta} = fc_{j\Delta}$, \identifyaffball refers to \bs in \fb to mark the balls with type code $tc_j$ as \affballsx, and resets $fc_{j\Delta}$ to $(1, \ldots, 1)$.
The condition $tc_j\& fc_{j\Delta} = fc_{j\Delta}$ holds as long as
(a) the $i$-th ($i\in [1, h]$) bit of $fc_{j\Delta}$ is 0, \ie there exists an edge/node deletion on ${P}_{fi}$, which may produce more matched nodes, or
(b) the $i$-th ($i\in [1, h]$) bit of $fc_{j\Delta}$ and $tc_j$ are both 1, \ie  balls with $tc_j$ already match with ${P}_{fi}$, though there are no edge/node deletions on ${P}_{fi}$.


\begin{example}
\label{exa-identifyaffball}
Consider the input and auxiliary structures in Example~\ref{exa-matchindex}, and \bfc in Fig.~\ref{fig-auxiliary-structures}.

\sstab(1) $\Delta P_1$ comes with a unit edge deletion $\delta_1$=$(\kw{SD},$ $\kw{ST})^-$ on $P_{f_2}$, and all four $fc$ in \bfc are updated from $fc$ (1, 1) to $fc_{\Delta}$ (1, 0), as shown in the second column of \bfc in Fig.~\ref{fig-auxiliary-structures}(b).
\identifyaffball identifies ball $\ball{[\kw{PM_1},2]}$ with $tc$ $(1, 1)$ and $\ball{[\kw{PM_2},2]}$ with $tc$ $(1, 0)$ as \affballsx,
	since $tc (1, 1) \& fc_{\Delta}(1, 0) = fc_{\Delta}(1, 0)$ and $tc (1, 0) \& fc_{\Delta} (1, 0) = fc_{\Delta}(1, 0)$.
Then \identifyaffball resets the two corresponding filtering codes in \bfc to $(1, 1)$.

\sstab(2) Consider another case when $\Delta P'_1$ comes with $\delta_1$ and $\delta_2$, where $\delta_1$ is same as above and $\delta_2 = (\kw{BA},\kw{UD})^+$. \bfc is updated as shown in the second column of \bfc in Fig.~\ref{fig-auxiliary-structures}(c), and \identifyaffball identifies the same \affballsx as above.
\end{example}


The correctness of \identifyaffball is ensured by the following. 

\begin{prop}
\label{prop-canaffballs}
For any ball $\ball{[v,r]}$ in $G$, if there exists a perfect subgraph of $P\oplus \Delta P$ in $\ball{[v,r]}$,
then $\ball{[v,r]}$ must be an affected ball produced by procedure \identifyaffball.
\end{prop}


\stitle{Lazy update policy}. To reduce computation, \incp only updates the partial relations for \affballsx in $\tilde{M}(P,G)$ for computing  $L_k(P\oplus \Delta P, G)$.
However, those partial relations in the filtered balls also need an update for handling future updates $\Delta P'$, but
definitely become outdated \wrt $P\oplus \Delta P$.
Hence, \incp needs a smart policy to maintain those match relations in the filtered balls.

To do this, algorithm \incp maintains the status of all unit updates applied to $P$ so far, and processes unit updates in $\Delta P$ as {\em late} as possible, while having no effects on future updates $\Delta P'$, \ie a {\em lazy update policy}.

Algorithm \incp utilizes auxiliary structure \upl together with the $cflag$ item in \bs.
When handling current $\Delta P$, for each ball $\ball{}$, $\ball{}[\kw{cflag}]$ records the id of the latest processed unit pattern update for $\ball{}$, and is initialized to $0$.
When future $\Delta P'$ comes, for any \affballx $\ball{}$ \wrt $\Delta P'$ and any fragment $P_{fi}$,
\incp computes $M(P_{fi}\oplus\Delta P'_{fi}, \ball{})$ based on $M(P_{fi}, \ball{})$ by procedure \incmatch (to be seen shortly),
where $\Delta P'_{fi}$ consists of the unit updates stored in $T(P_{fi})$ whose ids are larger than $\ball{}[\kw{cflag}]$ in \bs.

\begin{example}
\label{exa-lazyupdate}
Continue Example~\ref{exa-identifyaffball}.
(1) Balls $\ball{[\kw{PM_1},2]}$ and $\ball{[\kw{PM_2},2]}$ are \affballsx, and
\upl is shown in Fig.~\ref{fig-inc-maintain-unit}.

\sstab(a) \upl is updated \wrt\ $\Delta P_1=\{\delta_1\}$.
\incp updates the partial relations for $P_{f2}$ \wrt\ $\delta_1$ in the two balls,
and sets their $cflag$ in \bs to $\delta_1$, as the status shown in Fig.~\ref{fig-auxstr-index}.

\sstab
(b) Afterwards, $\Delta P_2$ with an edge insertion $\delta_2 = (\kw{BA},\kw{UD})^+$ comes.
\identifyaffball updates \bfc and \upl as shown in Fig.~\ref{fig-inc-maintain-unit} and identifies balls with $tc$ $(1, 1)$ as \affballsx, \eg $\ball{[\kw{PM_1},2]}$.

\sstab
(c) Finally, $\Delta P_3$ with a node deletion $\delta_3$ = $(\kw{UD})^-$ comes.
\identifyaffball identifies $tc$ $(1, 1)$, $(0, 1)$ and $(0, 0)$ as \affballsx.
Take ball $\ball{[\kw{BA_3}, 2]}$ for example, which is the first time identified as an \affballx.
By referring to \upl, \incp updates the partial relations for $P_{f1}$ \wrt\ $\{\delta_2, \delta_3\}$,
and for $P_{f2}$ \wrt\ $\{\delta_1\}$.
\looseness=-1

\sstab
(2) In the case when $\Delta P'_i$ contains multiple\eat{unit} updates, \bfc and \upl are updated accordingly as shown in Fig.~\ref{fig-auxiliary-structures}(c).
\end{example}


\stab{\bf (II) Updating Fragment-Ball matches}. We then update the partial match relations for \affballsx in $\tilde{M}(P,G)$ \wrt $\Delta P$,
by procedure \incmatch.

\stitle{Procedure \incmatch}. Given $h$-fragmentation ${\cal P}_{h}$ of $P$, $G$, $\tilde{M}(P,G)$, $\Delta P$, \upl and \affballsx \wrt $\Delta P$.
\incmatch updates $M(P_{fi}, \ball{})$ to $M(P_{fi}\oplus\Delta P_{fi}, \ball{})$ in $\tilde{M}(P,G)$ for each fragment $P_{fi}$ and each \affballx $\ball{}$.
Recall that $\Delta P_{fi}$ consists of unprocessed unit updates accumulated in \upl applied to $P_{fi}$.
We show how to update $M(P_{fi}, \ball{})$ in different cases.


\eetitle{(1) There exist edge/node deletions in $\Delta P_{fi}$}.
In this case, \incmatch accesses the \affballx $\ball{[v,r]}$ in $G$.
It simply computes the maximum match relations for $P_{fi}\oplus \Delta P_{fi}$ in $\ball{[v,r]}$
by procedure \rgraphsim in $O(|P_{fi}\oplus \Delta P_{fi}||\ball{[v,r]}|)$ time.

\eetitle{(2) No edge/node deletions in $\Delta P_{fi}$}.
\incmatch processes updates of the same type together in this case as follows.

\sstab{(i) Capacity changes in $\Delta P_{fi}$ or updates on $C$}.
In this case, no computation is needed for maintaining partial relations for \affballsx at all,
\ie $M(P_{fi}\oplus\Delta P_{fi}, \ball{}) = M(P_{fi}, \ball{})$.
Only a capacity check and an inner ball check in the combination procedure are needed (to be seen immediately).


\sstab{(ii) Edge insertions in $\Delta P_{fi}$}.
In this case, \incmatch calls procedure \patedgeinsert to process edge insertions.

\begin{figure}[t!]
\begin{center}
{\small
\myhrule
\vspace{-2ex}
\mat{0ex}{
\sstab {\sl Input:\/} $M(P_{fi}, \ball{[v,r]})$, $\ball{[v,r]}$, pattern edge insertion $\delta$ = $(u,u')^+$.\\
{\sl Output:\/} $M(P_{fi}\oplus \delta, \ball{[v,r]})$.\\
\sstab \bcc \hspace{1ex} \= $\kw{RMv}:=\emptyset$;\\
\icc \> \For \Each $u \in V_{P}$ \Do $\matchs(u)$ := \{$w | (u,w) \in M(P_{fi}, \ball{[v,r]})$\};\\
\icc \> \For \Each node $w \in \matchs(u)$\eat{in $M(P_{fi},\ball{[v,r]})$} \Do\\
\icc \>\hspace{2ex}\= \If there exists no $(w, w')\in E_{\ball{[v,r]}}$ with $w' \in \matchs(u')$ \Then\\
\icc \>\>\hspace{2ex}\= $\kw{RMv}.\kw{push}([u,w])$;\\
\icc \> \For \Each node $w' \in \matchs(u')$\eat{$M(P_{fi},\ball{[v,r]})$} \Do\\
\icc \>\> \If there exists no $(w', w)\in E_{\ball{[v,r]}}$ with $w \in \matchs(u)$ \Then\\
\icc \>\>\> $\kw{RMv}.\kw{push}([u',w'])$;\\
\icc \> \While $\kw{RMv}\neq \emptyset$ \Do\\
\icc \>\hspace{2ex}\= $[u,w]:=\kw{RMv}.\kw{pop}$(); $\matchs(u):=\matchs(u)\setminus\{w\}$;\\
\icc \>\> \For \Each $(u,u')\in E_{P_{fi}}$ \Do\\
\icc \>\>\hspace{2ex}\= \For \Each $(w, w')\in E_{\ball{[v,r]}}$ with $w' \in \matchs(u')$ \Do\\
\icc \>\>\>\hspace{2ex}\= \If there is no $(w', w'')\in E_{\ball{[v,r]}}$ with $w'' \in \matchs(u)$ \Then\\
\icc \>\>\>\>\hspace{2ex}\= \kw{RMv}.\kw{push}([$u',w'$]);\\
\icc \> \If there is a node $u\in V_{P_{fi}}$ with $|\matchs(u)| = 0$ \Then $\matchs(\cdot):=\emptyset$;\\
\icc \> $M(P_{fi}\oplus \delta, \ball{[v,r]})$ := \{$(u,w) | u \in V_{P}, w \in \matchs(u)$\};\\
\icc \> \Return $M(P_{fi}\oplus \delta, \ball{[v,r]})$;
}
\vspace{-2.5ex} \myhrule
}
\end{center}
\vspace{-3ex}
\caption{Procedure \patedgeinsert} \label{alg-patedgeinsert}
\vspace{-4ex}
\end{figure}

\stitle{Procedure \patedgeinsert}.
Given $M(P_{fi}, \ball{[v,r]})$ (also represented by $\matchs(\cdot)$), $\ball{[v,r]}$ and an edge insertion $\delta=(u,u')$, \patedgeinsert computes $M(P_{fi}\oplus\delta, \ball{[v,r]})$ {\em incrementally}, as shown in Fig.~\ref{alg-patedgeinsert},
along the same lines as for data incremental graph simulation~\cite{FanWW13-tods}.
\patedgeinsert first finds the directly affected data nodes that need to be removed from $\matchs(\cdot)$ due to the edge insertion to $P_{fi}$,
and pushes them along with the matched pattern nodes into $\kw{RMv}$ (lines 3-8).
It then recursively identifies and removes the nodes in $\matchs(\cdot)$ affected by the previous removed nodes (lines 9-14).
The recursive process is executed by utilizing a stack $\kw{RMv}$.
If there exists a pattern node $u$ with empty $\matchs(u)$, then $\matchs(\cdot)$ is set to $\emptyset$ (line 15).
Finally, \patedgeinsert returns the updated $\matchs(\cdot)$ for $P_{fi}\oplus \delta$ (lines 16-17).
\looseness=-1

\begin{example}
\label{exa-edge-insertion}
Consider case (1)-(b) in Example~\ref{exa-lazyupdate}.
Given $\delta_2 = (\kw{BA},\kw{UD})^+$, and $M(P_{f1}, \ball{}_{\kw{PM_1}})$, which is composed of nodes \kw{PM_1}, \kw{BA_1} and \{\kw{UD_1},\kw{UD_2}\} mapped to nodes \kw{PM}, \kw{BA} and \kw{UD} in $P_{f1}$.
To compute the updated $M(P_{f1}\oplus \delta_2, \ball{}_{\kw{PM_1}})$,
\patedgeinsert removes \kw{UD_2} that is directly affected by $\delta_2$, and finds no other nodes need to be removed.
\end{example}


\sstab{(iii) Node insertions in $\Delta P_{fi}$}.
Node insertions are handled in a similar way as edge insertions, by extending \patedgeinsert.

Given a node insertion $\delta$ = $(u,(u,u'))^+$, where $u$ is a newly inserted node,
to compute the updated $M(P_{fi}\oplus \delta, \ball{[v,r]})$,
\incmatch firstly computes the set of nodes $\matchs(u)$ in $\ball{[v,r]}$ that have the same label with $u$, and then calls \patedgeinsert($M(P_{fi}, \ball{[v,r]})$, $\ball{[v,r]}$, $(u,u')$) to get the updated result.

\vspace{0.5ex}
\stitle{Updating \fb.} After updating $\tilde{M}(P,G)$,
\incp updates \fb for all \affballsx, by changing the links according to the updated partial relations in $\tilde{M}(P,G)$, and also updating the \kw{cflag} item in \bs, which is in $O(||\affballsx||)$ time.


\stab{\bf (III) Combining Fragment-Ball matches}. Algorithm \incp finally combines the updated partial match relations in \affballsx to get the updated top-$k$ perfect subgraphs $L_{k}(P\oplus \Delta P,G)$ by procedure \comb. Observe that only the balls from \affballsx that match with all pattern fragments of $P\oplus \Delta P$ can enter the combination process.

\stitle{Procedure \comb}. For an \affballx $\ball{[v,r]}$, \comb invokes $\patedgeinsert(\bigcup_{i\in [1, h]}M(P_{fi},\ball{[v,r]}), \ball{[v,r]}, C\oplus \Delta C)$
to compute the maximum match relations of $P\oplus \Delta P$ for $\ball{[v,r]}$ incrementally,
where $\Delta C$ consists of the edge insertions/deletions in $\Delta P$ applied to the cut edges $C$.
It then checks whether the capacity bounds, together with the updates on them, are satisfied.
If so, it constructs the perfect subgraph \wrt the match relations above.
It then checks the inner balls together with the capacity bounds, and finally returns the list of top-$k$ perfect subgraphs $L_{k}(P\oplus \Delta P,G)$.

\begin{example}
\label{exa-combination}
Continue Example~\ref{exa-lazyupdate}-(1), after \incp updated partial relations for \affballsx \wrt $\Delta P_3$,
balls $\ball{[\kw{PM_1},2]}$, $\ball{[\kw{PM_2},2]}$ and $\ball{[\kw{BA_3},2]}$ enter the combination process.

\sstab
(1) For $\ball{[\kw{PM_1},2]}$ and $\ball{[\kw{PM_2},2]}$, as $C\oplus \Delta C=\{(\kw{PM},\kw{SA})\}$,
\comb finds that there is an \kw{SA_i} (resp.\,\kw{PM_j}) connecting to \kw{PM_j} (resp.\,\kw{SA_i}),
and the capacity bounds are satisfied.
For inner balls, based on above results, \comb finds that no perfect subgraphs reside in $\ball{[\kw{PM_1},1]}$ and $\ball{[\kw{PM_2},1]}$.
Hence it returns the above two perfect subgraphs in two balls.

\sstab
(2) For $\ball{[\kw{BA_3},2]}$, \comb finds no \kw{SA_i} connecting to \kw{PM_j}, and vice versa.
Hence, no  sensible matches are found.
\end{example}


\stab{\bf (IV) Early return optimization technique}.
We propose an optimization technique for \incp to further speed-up the incremental computations, by making use of the top-$k$ semantics.
We first define {\em early return} for incremental top-$k$ algorithms, analogous to {\em early termination} for batch top-$k$ algorithms~\cite{FaginLotem03}.

\stitle{Early return}. An algorithm has the {\em early return property},
if for pattern $P$ with updates $\Delta P$ and for any data graph $G$,
it outputs $L_k(P\oplus \Delta P, G)$ as early as possible without the need to update match relations for every \affballx,
while the updates can be executed in background.

\begin{prop}
\label{prop-early-return}
There exists an algorithm for the dynamic top-$k$ team formation problem with early return property.
\end{prop}

We prove \incp retains the early return property.
Recall the density based filtering optimization for algorithm \optgrouprec in Section~\ref{sec-tsimAlg}.
\incp also utilizes density upper bounds for pruning a portion of \affballsx.
More specifically, given $P$ and $G$, \incp maintains the density upper bound for each ball in the $den$ item in \bs, \ie $\ball{}[\kw{den}]$,
calculated according to Lemma~\ref{lemma-approximation-bound}.
Thus, given $\Delta P$, if the top-$k$ densest perfect subgraphs found so far are denser than the density upper bound of the remaining \affballsx,
\incp\ {\em outputs} the current top-$k$ densest perfect subgraphs as $L_k(P\oplus \Delta P, G)$,
while continuing updating $\tilde{M}(P, G)$ in \affballsx in {\em background}.

Note that the early return optimization is  effective for pattern updates, but not for data updates and the case when $\Delta P$ contains node insertions with new labels (expertise).

\begin{figure}[t!]
	\begin{center}
		{\small
			\myhrule \vspace{-2ex}
			\mat{0ex}{
				\sstab {\sl Input:\/} \= $P$, $h$-fragmentation ${\cal P}_{h}$, $G$, integers $r$ and $k$, $\Delta P$, and\\
				\>auxiliary structures $\tilde{M}(P,G)$, \fb, \bfc and \upl.\\
				{\sl Output:\/} Top-$k$ perfect subgraphs for $P \oplus \Delta P$ in $G$.\\
				\sstab \bcc \hspace{1ex} \= $L_{k}$ := $\emptyset$;\\
				\icc \> \affballsx\ :=  $\identifyaffball({\cal P}_h, \Delta P, \fb, \bfc)$;\\
				\icc \> Sort \affballsx by $\ball{}[\kw{den}]$ in non-ascending order;\\
				\icc \> \For \Each $\ball{[v,r]}$ in \affballsx \Do\quad/* non-ascending order */\\
				\icc \> \hspace{3ex} \= \If $|L_{k}| \geq k$ \And $\ball{[v,r]}[\kw{den}] \leq \density{L_{k}[k-1]}$ \Then \\
				\icc \>\>\hspace{2ex} \= \bf{Output} $L_{k}[0:k-1]$. /* early-return optimization*/\\
				\icc \>\>$\incmatch(M(P_{fi},\ball{[v,r]}),\ball{[v,r]}, \Delta P_{fi})$ $(i\in[1,h])$;\\
				\>\> /* runs in the background */\\
				\icc \>\> $S_{G_s} := \comb(\bigcup_{i\in [1, h]}M(P_{fi},\ball{[v,r]}), \ball{[v,r]}, C\oplus \Delta C)$;\\
				\icc \>\> Insert the set of perfect subgraphs in $S_{G_s}$ into $L_{k}$;\\
				\icc \> \Return $L_{k}[0:k-1]$.
			}
			\vspace{-2.5ex} \myhrule
		}
	\end{center}
	\vspace{-3ex}
	\caption{Algorithm \incp} \label{alg-optInc}
	\vspace{-4ex}
\end{figure}


\stab{\bf (V) The complete algorithm for pattern updates}.
Given $\tilde{M}(P,G)$, \fb, \bfc and \upl, for pattern update $\Delta P$,
algorithm \incp computes the top-k perfect subgraphs for $P \oplus \Delta P$ in $G$ with early return property, and maintains auxiliary structures simultaneously by invoking procedure \identifyaffball, \incmatch and \comb one by one.

\stitle{Algorithm \incp}. It works as follows, as shown in Fig.~\ref{alg-optInc}.
For each $\Delta P$, \incp firstly sets the result list $L_{k}$ to empty,
and identifies \affballsx \wrt $\Delta P$ by \identifyaffball (lines 1-2).
It then sorts \affballsx by their density upper bounds $\ball{}[\kw{den}]$ in \bs in non-ascending order (line 3),
and accesses \affballsx sequentially by this order (lines 4-10).
Whenever it comes to next \affballx $\ball{[v,r]}$, it firstly checks whether there are already $k$ perfect subgraphs found in $L_{k}$, and moreover,
the density of the $k$th (smallest) perfect subgraph in $L_{k}$ is larger than $\ball{[v,r]}[\kw{den}]$ (line 5).
If so, \incp immediately outputs $L_{k}$ as final results (line 6),
and then continues to update $\tilde{M}(P,G)$ for those \affballsx in background by \incmatch (line 7);
Otherwise, \incp updates the partial relations and combines them by \comb to get the set of perfect subgraphs $S_{G_s}$ in $\ball{[v,r]}$ and its inner balls (lines 7-8).
It then inserts the set of perfect subgraphs in $S_{G_s}$ into $L_{k}$ (line 9).

\stitle{Correctness \& complexity analysis}.
The correctness of \incp is assured by the correctness of \identifyaffball (Proposition~\ref{prop-canaffballs}), \incmatch, \comb, and early return property (Lemma~\ref{lemma-approximation-bound}).
\incp is in $O(\bigcup_{\hat{G}\in \affballsx}\bigcup_{i\in[1,h]}(|M(P_{fi}, \hat{G})|$ + $r|M(P_{fi} \oplus \Delta P_{fi}, \hat{G})|$) + $r|P \oplus \Delta P||\affballsx|$+$|\Delta P|)$ time \wrt $\Delta P$, while $r$ is small, \ie 2 or 3 (See full version~\cite{fullvldb18}).

\subsection{Dealing with Data Updates}
\label{subsec-Ginc}

We next propose \incd to handle $\Delta G$, following the framework in Section~\ref{subsec-framework}.
Given auxiliary structures $\tilde{M}(P,G)$ and \fb,
we put together procedures \identifyaffball, \incmatch and \comb for computing match results for $P$ in $G \oplus \Delta G$.
As procedures \incmatch and \comb handle $\Delta G$ basically the same as $\Delta P$, so we mainly show how \identifyaffball identifies \affballsx \wrt $\Delta G$.

\stitle{Procedure \identifyaffball}.
Given $G$, $\Delta G$, and \fb, \identifyaffball identifies \affballsx according to the lemma as follows.

\begin{lemma}
\label{lemma-incgrdata-affballs}
A ball $\ball{[v,r]}$ with center node $v$ in $G$ is identified as an \affballx \wrt $\Delta G$ and \fb,

\sstab{(1)} for some unit data update $\delta$ of $\Delta G$, where
(a) $\delta$ is an edge insertion/deletion, $(w_1,w_2)^+/(w_1,w_2)^-$ and $v$ is in  both $\ball{[w_1,r]}$  and $\ball{[w_2,r]}$, or
(b) $\delta$ is a node insertion/deletion, $(w,(w,w'))^+$/$(w)^-$ and $v$ is in $\ball{[w,r]}$; or

\sstab{(2)} when $\ball{[v,r]}$ has type code $(1,\ldots,1)$  in \fb.
\end{lemma}

We say a ball which satisfies condition (1) is a {\em structural affected} ball,
\ie the structure of the ball is changed due to the exertion of some updates in $\Delta G$.

\vspace{-0.5ex}
\begin{prop}
\label{prop-affected-datainc}
Given $P$, $G$ and $\Delta G$, if there is a perfect subgraph for $P$ in  ball $\widehat{G\oplus\Delta G}{[v,r]}$ of $G\oplus \Delta G$,
then $\ball{[v,r]}$ must be an affected ball produced by procedure \identifyaffball.
\end{prop}


Different from pattern updates, procedure \incmatch recomputes partial match relations in $\tilde{M}(P,G)$
for each pattern fragment of $P$ in each {\em structural affected} ball;
and no computation is needed for \affballsx that only satisfy condition (2) in Lemma~\ref{lemma-incgrdata-affballs}.
Procedure \comb combines the partial relations \wrt $\Delta G$ in the same way as handling $\Delta P$.

\stitle{Updating \fb}. Algorithm \incd also updates \fb for all \affballsx.
In addition to updating the links from \fs to \bs in \fb as for pattern updates,
\incd maintains \bs by
(a) removing (resp. inserting new) entries from (resp. to) \bs corresponding to balls whose center nodes are removed from (resp. inserted to) $G$, due to node deletions (resp. node insertions);
and (b) updating the \kw{den} item in \bs \wrt $\Delta G$.
These updates can be done in $O(|\affballsx|)$ time.

\begin{example}
\label{exa-Ginc}
Consider $P_1$ and $G_1$ (both without dashed edges) in Fig.~\ref{fig-motivation-example}, and \fb in Fig.~\ref{fig-auxiliary-structures}(a).
When $\Delta G_1=(\kw{SD_3},\kw{ST_3})^+$ comes,
by Lemma~\ref{lemma-incgrdata-affballs},
\identifyaffball identifies $\ball{[\kw{SD_3}, 2]}$, $\ball{[\kw{ST_3},2]}$, $\ball{[\kw{SA_3}, 2]}$ and $\ball{[\kw{PM_2}, 2]}$ as {\em structural affected} balls,
together with balls with $tc$ $(1, 1)$ in \fb as \affballsx, \ie $\ball{[\kw{PM_1}, 2]}$, while filtering out all other balls.
\end{example}

\stitle{Algorithm \incd.} Given $\tilde{M}(P,G)$, \fb and data updates $\Delta G$,
\incd computes the match results for $P$ in $G \oplus \Delta G$, and maintains
auxiliary structures by invoking procedures \identifyaffball, \incmatch and \comb sequentially.

\stitle{Correctness \& complexity analyses}. The correctness of \incd \wrt $\Delta G$ follows from Proposition~\ref{prop-affected-datainc} and the correctness of \incmatch and \comb.
\incd is overall in
$O(\bigcup_{\widehat{G\oplus\Delta G}\in \affballsx}\bigcup_{i\in[1,h]}$ $r|M(P_{fi}, \widehat{G\oplus\Delta G})|$+$r|P||\affballsx| + |\Delta G|)$ time \wrt $\Delta G$, while $r$ is small, \ie 2 or 3  (See~\cite{fullvldb18}).
\looseness=-1

\subsection{Unifying Pattern and Data Updates}
\label{subsec-completePG}

We are now ready to provide algorithm \inc, integrating \incp and \incd, presented in Sections~\ref{subsec-Qinc} and~\ref{subsec-Ginc}, respectively, to process continuous pattern and data updates, separately and simultaneously.

Algorithm \inc is able to handle {\em simultaneous} $\Delta P$ and $\Delta G$, because of the consistency in:
(1) the processes for handling $\Delta P$ and $\Delta G$, which follow the same steps in Section~\ref{subsec-framework};
(2) auxiliary data structures for supporting $\Delta P$ and $\Delta G$; and
(3) the combination procedures, which suffice to support simultaneous pattern and data updates.

Observe that \inc can handle {\em continuously} simultaneous $\Delta P$ and $\Delta G$,
as \inc incrementally maintains the auxiliary structures for continuously coming $\Delta P$ and $\Delta G$.

\stitle{Remark}. Note that the running time of \incp and \incd is determined by \{$P, \Delta P, \tilde{M}(P,G)$, $\affballsx$\}
and \{$P, \Delta G, \tilde{M}(P,G)$, $\affballsx$\}, respectively, not directly depending on $G$.
From this, we complete the proof of Theorem~\ref{thm-framework-inc}.

\balance
\section{Experimental Study}
\label{sec-expt}

\begin{figure*}[tb!]
\begin{center}
	
\subfigure[{\scriptsize Varying $|V_{P}|$ (\citationd)}]{\label{fig-exp-semantic-citation-diameter}
\includegraphics[scale=0.38]{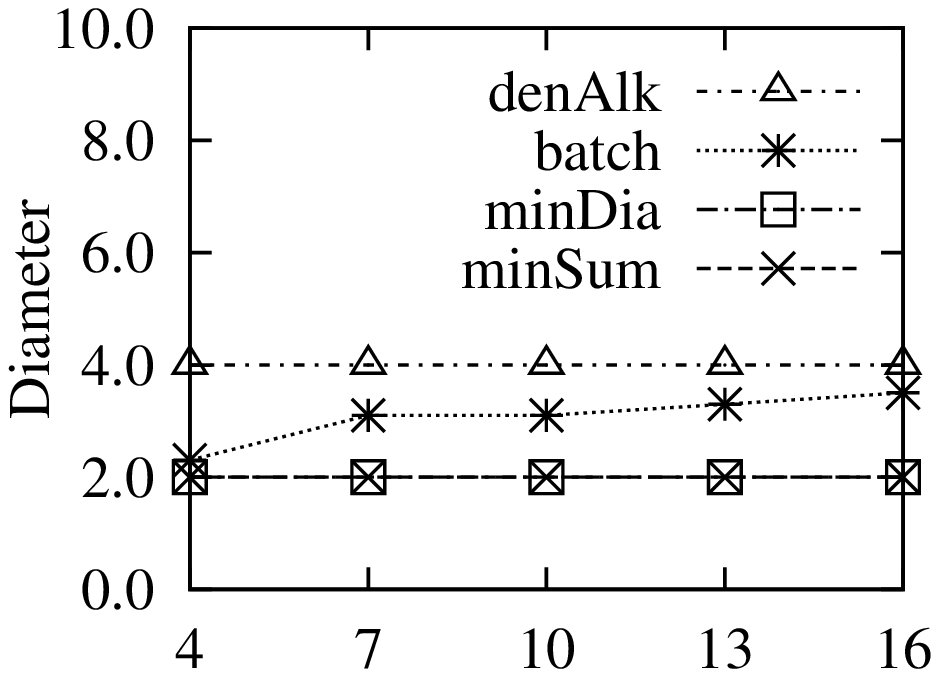}}
\hspace{0.2ex}
\subfigure[{\scriptsize Varying $|V_{P}|$ (\citationd)}]{\label{fig-exp-semantic-citation-density}
\includegraphics[scale=0.38]{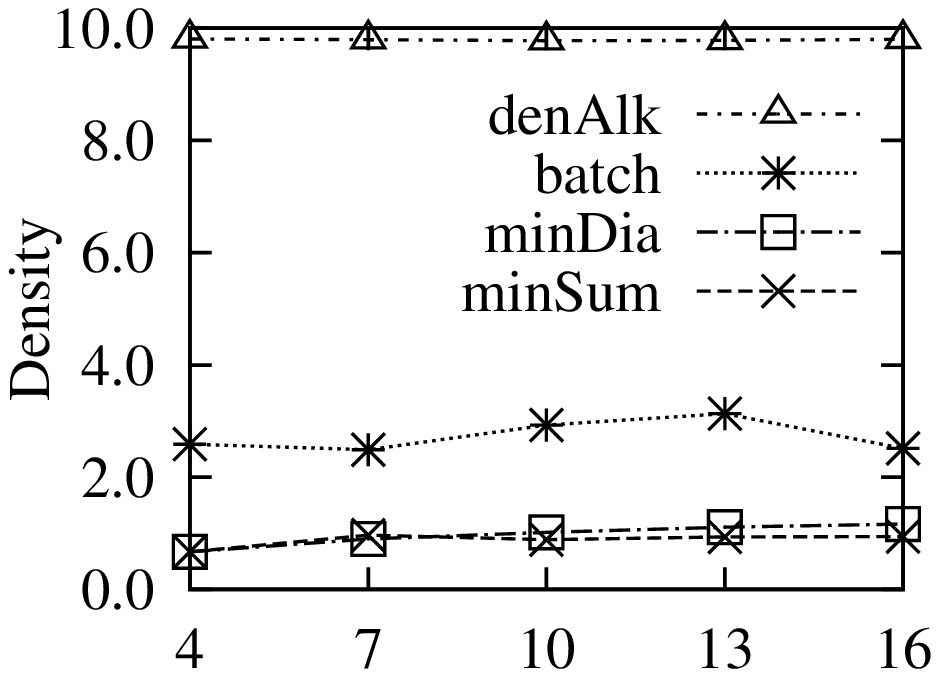}}
\hspace{0.2ex}
\subfigure[{\scriptsize Varying $|V_{P}|$ (\citationd)}]{\label{fig-exp-semantic-citation-capacity}
\includegraphics[scale=0.38]{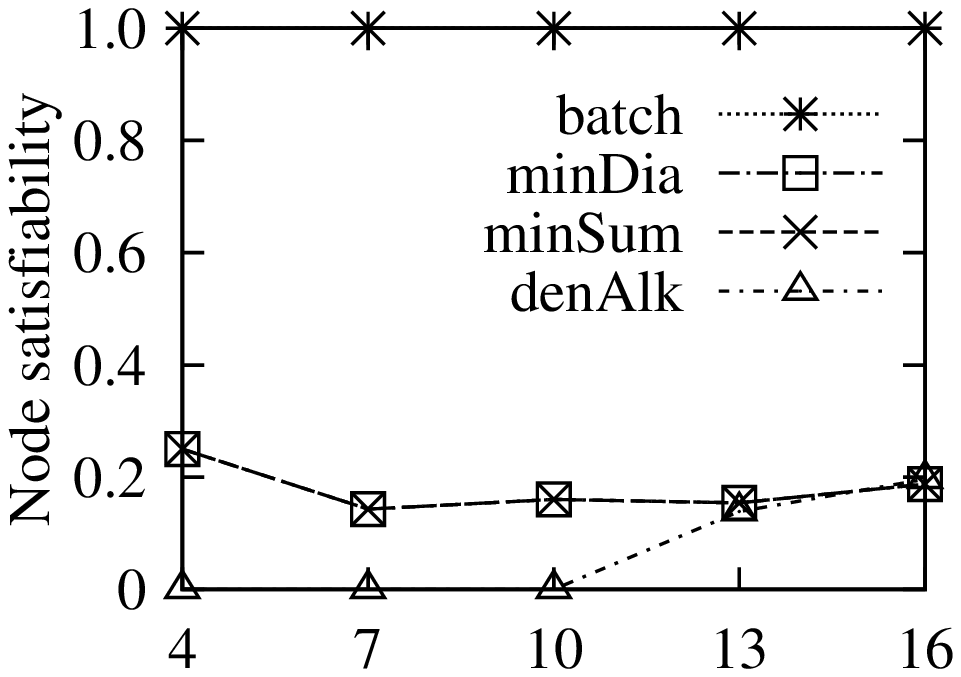}}
\hspace{0.2ex}
\subfigure[{\scriptsize Varying $|V_{P}|$ (\citationd)}]{\label{fig-exp-semantic-citation-link}
\includegraphics[scale=0.38]{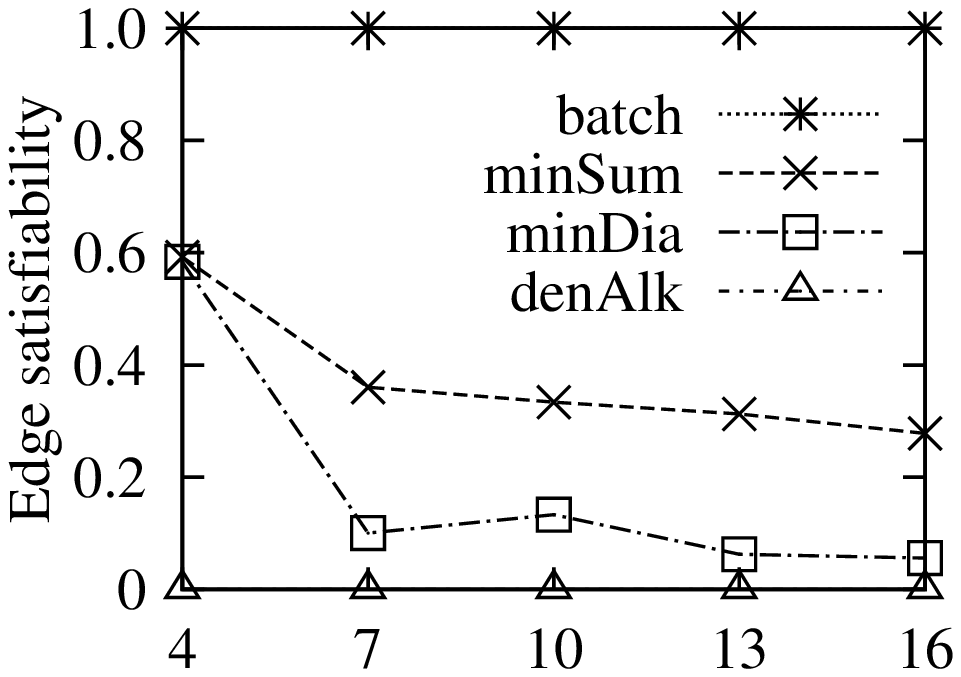}}
\vspace{-2.5ex}

\subfigure[{\scriptsize Varying $k$ (\citationd)}]{\label{fig-exp-semantic-citation-diameter-varyk}
\includegraphics[scale=0.38]{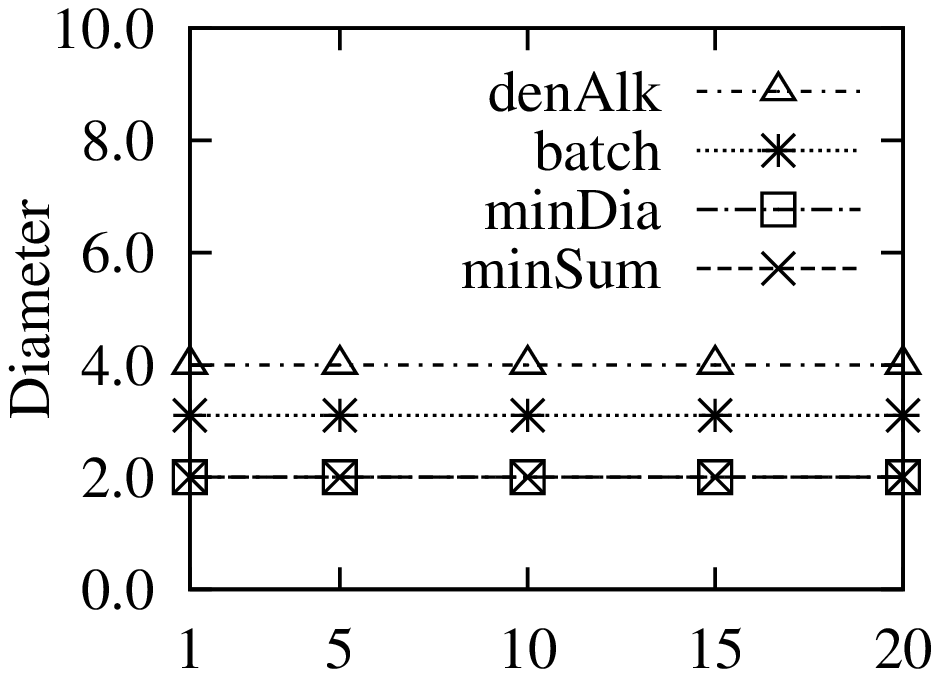}}
\hspace{0.2ex}
\subfigure[{\scriptsize Varying $k$ (\citationd)}]{\label{fig-exp-semantic-citation-density-varyk}
\includegraphics[scale=0.38]{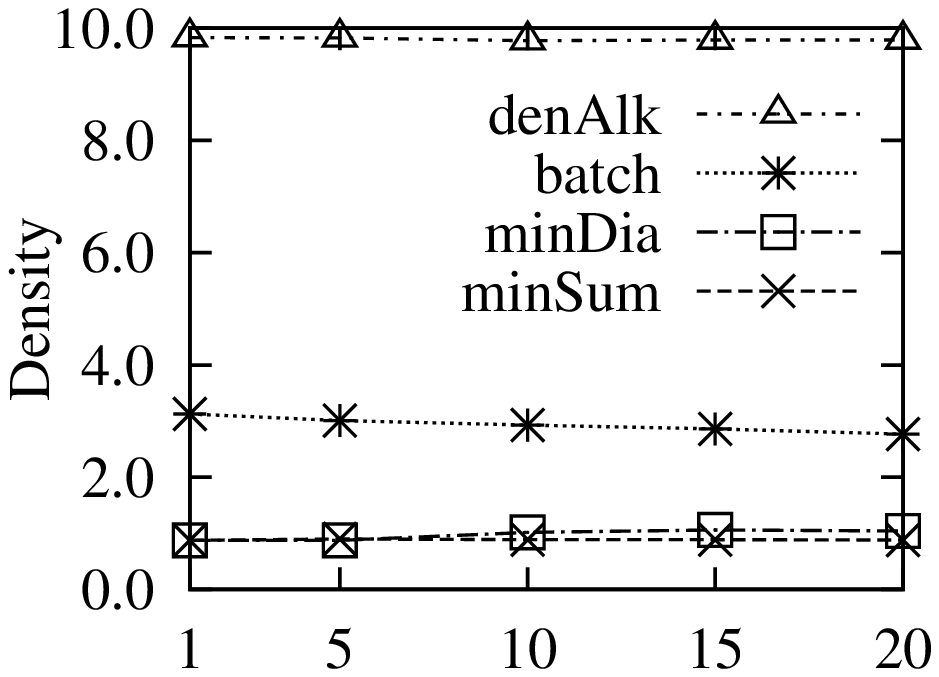}}
\hspace{0.2ex}
\subfigure[{\scriptsize Varying $k$ (\citationd)}]{\label{fig-exp-semantic-citation-capacity-varyk}
\includegraphics[scale=0.38]{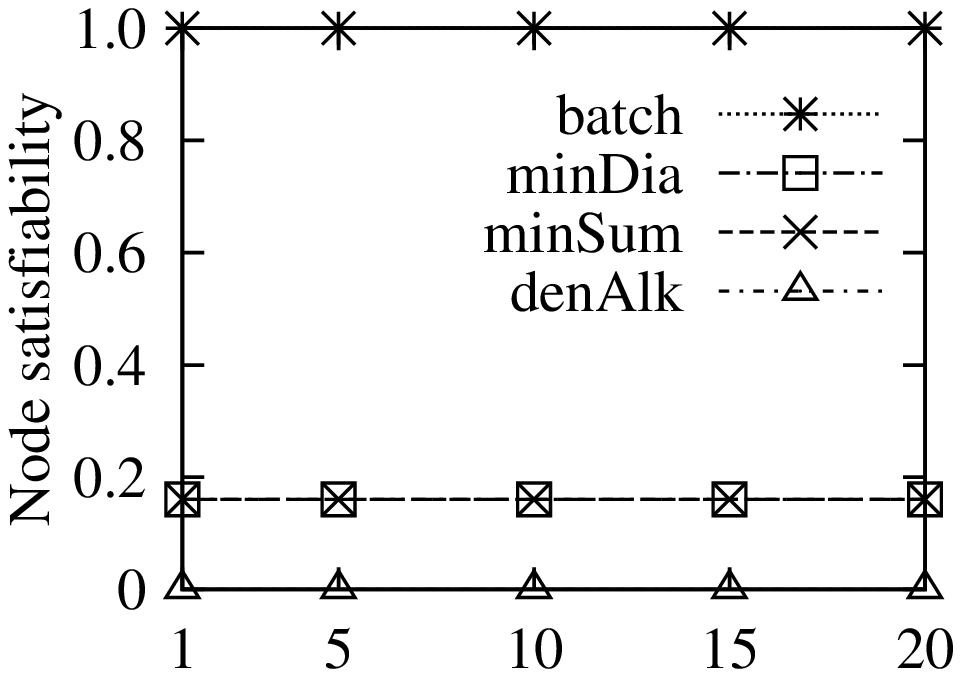}}
\hspace{0.2ex}
\subfigure[{\scriptsize Varying $k$ (\citationd)}]{\label{fig-exp-semantic-citation-link-varyk}
\includegraphics[scale=0.38]{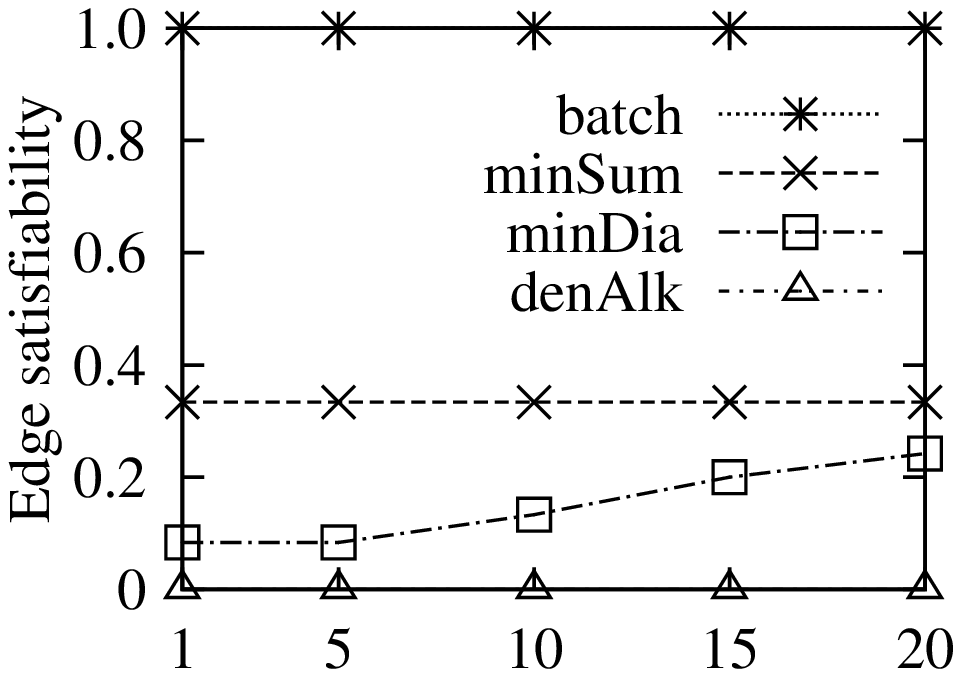}}
\vspace{-2.0ex}
\end{center}

\vspace{-3ex}
\caption{Performance evaluation of algorithm \optgrouprec for top-$k$ team formation}
\label{exp-semantic-effectiveness-citation}
\vspace{-3.0ex}
\end{figure*}

We conducted four sets of experiments to evaluate the performance of
(1) \optgrouprec for the top-$k$ team formation problem,
(2) \inc for the dynamic top-$k$ team formation problem
\wrt single sets of (a) pattern updates, (b) data updates, and (c) simultaneous pattern and data updates,
(3) \inc \wrt continuous sets of pattern and data updates, and
(4) the extra space cost of auxiliary structures used by \inc.

\subsection{Experimental Settings}

 We used the following real-life and synthetic datasets.


\vspace{-0.3ex}
\etitle{Real-life graphs.} We used two real-life graphs.

\ni (1) {\em \citationd}~\cite{citationWeb} contains 1.39M paper nodes and 3.02M paper-paper citation edges.
We used its undirected version, where edges indicate the relevance relationship, and generated 200 labels based on phrase clustering of paper titles.

\ni (2) {\em \youtube}~\cite{youtubeWeb} contains 2.03M video nodes and 12.2M edges, which represent recommendations between two videos. We used the undirected version, and generated 400 labels based on the built-in categories and ages of videos.

\etitle{Synthetic graph generator.} We generated synthetic graphs (\synthetic) with community structures as existed in real-life, by adopting the LFR-benchmark graph model~\cite{AndreaSF08}. It is controlled by three parameters: the number $n$ of nodes, the average degree $d$ of nodes, and the number $l$ of node labels.

\eat{
 We used the following datasets and settings.	
	
\etitle{Data graphs.} We used two real-life and a synthetic datasets.

\ni (1) {\em \citationd}~\cite{citationWeb} is a real-life dataset containing 1.39M paper nodes and 3.02M paper-paper citation edges.
We used its undirected version, where edges indicate the relevance relationship. We generated 200 node labels based on phrase clustering of paper titles.

\ni (2) {\em \youtube}~\cite{youtubeWeb} contains 2.03M video nodes and 12.2M edges, which represent recommendations between two videos. We used the undirected version, and generated 400 labels based on the built-in categories and ages of videos.

\ni (3) We generated synthetic data graphs (\synthetic) with community structures as existed in real-life networks, by adopting the LFR-benchmark graph model~\cite{AndreaSF08}. It is controlled by three parameters: the number $n$ of nodes, the average degree $d$ of nodes, and the number $l$ of node labels.

}

\etitle{Pattern generator}.
We implemented a generator to produce pattern graphs, controlled by 4 parameters:
the number of nodes $|V_P|$, the number of edges $|E_P|$, the label $l_{P}$ for each node from
an alphabet of labels in the corresponding data graphs,
and the capacity bound $f_{P}$ for each node.

\etitle{Algorithms}. We implemented the following algorithms, all in C++:
(1) algorithm \optgrouprec for \teamF, (2) incremental algorithm \inc for \dynteamF,
(3) three compared top-$k$ team formation algorithms \mindia, \minsumdis and \denalk, where
(a) \mindia is to minimize the team diameter~\cite{Lappas09}, which is firstly proposed for the team formation problem,
(b) \minsumdis is to minimize the sum of all-pair shortest distances of teams~\cite{Kargar11}, and
(c) \denalk is to maximize the team density~\cite{GajewarS12}, which has the same goal with our algorithms.
Most of the algorithms for \teamF, including  \mindia and \denalk, only compute the best team,
while \minsumdis is able to find top-$k$ teams in polynomial time, which is an adaption of Lawler's procedure \cite{Lawler1972}.
Based on this, we extend \mindia and \denalk to find top-$k$ teams in polynomial time.

We used a PC with Intel Core i5-4570 CPU and 16GB of memory. We randomly generated 3 sets of input and repeated 5 times for each test. The average is reported here.

\subsection{Experimental Results}

We present our findings. In all the experiments, we set $k=10$, $r=2$, $h=3$, $(|V_{P}|,$ $|E_{P}|)$ to be (10,12),
and capacity bounds to be [1,10] by default.
When generating synthetic graphs, we fixed $n=10^7$, $d=10$ and $l=200$.
All the findings on \youtube are reported in the full version~\cite{fullvldb18}.

\stitle{Exp-1: Efficiency of \optgrouprec}. We firstly evaluated the efficiency of \optgrouprec vs. \mindia, \minsumdis and \denalk.
We generated pattern graphs for \optgrouprec, and the corresponding queries (labels requirements) for \mindia, \minsumdis and \denalk.

Algorithms \mindia, \minsumdis and \denalk do not scale well on large graphs.
Indeed,  (1) \mindia and \minsumdis took more than $8$ hours to finish their preprocessing, \ie computing all-pair-shortest-paths;
and (2) \denalk took more than 24 hours even when $k=1$ on \citationd.
By contrast, \optgrouprec took around 100 seconds on \citationd by default settings.
Hence, we report the effectiveness of these algorithms on a sampled data graph with $10,000$ nodes on \citationd only.

\stitle{Exp-2: Effectiveness of \optgrouprec}. We then evaluated the efficiency of \optgrouprec vs. \mindia, \minsumdis and \denalk by checking the quality of matches returned by them.

\eat{
\etitle{(1) Case study.}
We find \optgrouprec is able to identify sensible matches meeting the practical requirements.
We illustrate this with a real-life example pattern graph.

As shown in Fig.~\ref{fig-exp-1-effect-citation}, pattern graph $P_C$ is to find
(a) a set of papers in \citationd that ``Databases'', ``Data Mining'' and ``Artificial Intelligence'' papers are related to each other;
(b) ``Data Mining'' papers are related to ``Software Engineering'' papers; and
(c) the required number of papers for each domain is shown on $P_C$.
In \citationd, nodes are papers with unique id from different domains,
with labels indicating their domains, and they only match the nodes of $P_C$
with the same geometry shapes, \ie circles, ellipses and squares.

The match result of $P_C$ is shown in Figure~\ref{fig-exp-1-effect-citation}.
Observe that subgraph $G_C$ is the top-1 match result found by \optgrouprec for $P_C$, but cannot be found by \mindia, \minsumdis and \denalk, while $G'_C$ and $G''_C$ are the top-1 results found by \mindia and \minsumdis respectively.
The top-1 result found by \denalk is the densest component composed of 819 nodes (omitted).
These tell us that \optgrouprec has more expressive power than the other team formation algorithms
by identifying sensible matches satisfying the structural and capacity requirements.}

To evaluate the quality of teams found by the above four algorithms for \teamF, we defined four quality measures.
Consider a matched subgraph $G_S$ and pattern $P(V_P, E_P)$.

\noindent {(a) [{\em Diameter}]}: the diameter of $G_S$.

\noindent {(b) [{\em Density}]}: the density of $G_S$.

\noindent {(c) [{\em Node satisfiability}]}: $\eta_V(G_S, P)$ = $\#\kw{sat}_V(G_S, P) / |V_P|$, where $\#\kw{sat}_V(G_S, P)$ is the number of nodes in $P$ that are satisfied by $G_S$, in which we say a pattern node $u$ is satisfied by $G_S$ if there are a set $V_u$ of nodes in $G_S$ that match $u$ and moreover, $V_u$ satisfies the capacity constraints on $u$.

\noindent {(d) [{\em Edge satisfiability}]}: $\eta_E(G_S, P)$ = $\#\kw{sat}_E(G_S,P)/|E_P|$, where $\#\kw{sat}_E(G_S,P)$ is the number of edges in $P$ satisfied by $G_S$, in which we say an edge $(u_1, u_2)$ is satisfied by $G_S$ if for each $v_1$ in $G_S$ that matches $u_1$,
there exists $(v_1, v_1')$ in $G_S$ so that $v_1'$ matches $u_2$, and for each  $v_2$ in $G_S$ that matches $u_2$, there exists $(v_2, v_2')$ in $G_S$ such that $v_2'$ matches $u_1$.
\looseness=-1

Note that (a) and (b) are two traditional quality measures utilized by existing team formation algorithms~\cite{GajewarS12,realTeamForm13,Lappas09,Kargar11}.
Intuitively, $\eta_V(G_S, P)$ (resp. $\eta_E(G_S,P)$) measures how well $G_S$ meets the node capacity requirements (resp. structural constraints) in $P$, and their values fall in $[0, 1]$.

\etitle{(i) Impacts of $|V_{P}|$}. Varying the number $|V_{P}|$ of
nodes in $P$ from 4 to 16, we took the average value of top-$k$ teams found by  \optgrouprec, \mindia, \minsumdis and \denalk \wrt four quality measures.
The results are reported in Figures \ref{fig-exp-semantic-citation-diameter}-\ref{fig-exp-semantic-citation-link}.

Observe the following.
(1) The diameters of teams found by \optgrouprec are comparable to those of \mindia and \minsumdis, which are in particularly designed to minimize the diameters, as shown in Fig.~\ref{fig-exp-semantic-citation-diameter}. This is ensured by the use of balls in \optgrouprec.
(2) The densities of teams found by \optgrouprec, though are smaller than \denalk which is specialized for maximizing team densities, are larger than \mindia and \minsumdis, as in Fig.~\ref{fig-exp-semantic-citation-density}.
(3) The node satisfiability of teams found by \optgrouprec is much higher than \mindia, \minsumdis and \denalk, \eg 1.0 vs. no larger than 0.2 in all cases as in Fig.~\ref{fig-exp-semantic-citation-capacity}.
(4) The teams found by \optgrouprec come with a higher edge satisfiability , \eg 1.0 in all cases, compared to smaller than 0.6 by \mindia, \minsumdis and \denalk, as shown in Fig.~\ref{fig-exp-semantic-citation-link}.


\etitle{(ii) Impacts of $k$}. Varying $k$ from 1 to 20, we report the results in Figures
\ref{fig-exp-semantic-citation-diameter-varyk}-\ref{fig-exp-semantic-citation-link-varyk}.
Observe that the quality of teams found by the four algorithms shows the same rule as varying $|V_{P}|$, and the quality is not sensitive to $k$, a desirable property when top-$k$ semantics is concerned.

These verify that \optgrouprec can effectively preserve structural and capacity constraints for top-$k$ team formation \wrt edge and node satisfiability, and pertains a good team collaboration compatibility \wrt diameter and density.

\begin{figure*}[tb!]
\begin{center}
\subfigure[{\scriptsize Varying $|\Delta P|$ (deletions)}]{\label{fig-exp-patinc-del}
\includegraphics[scale=0.38]{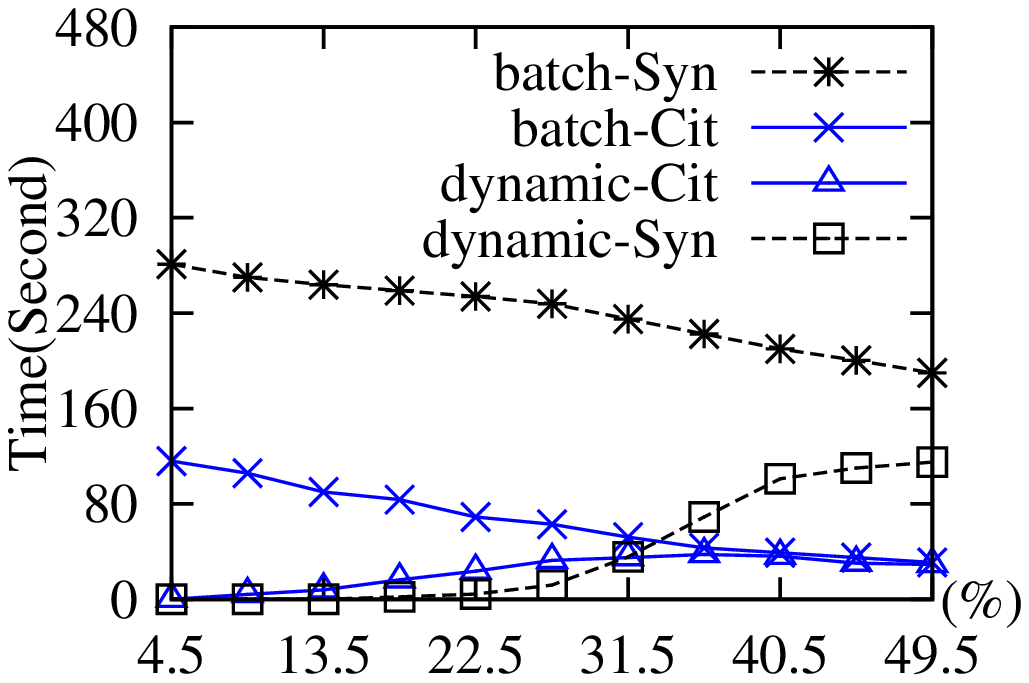}}
\hspace{0.2ex}
\subfigure[{\scriptsize Varying $|\Delta P|$ (insertions)}]{\label{fig-exp-patinc-ins}
\includegraphics[scale=0.38]{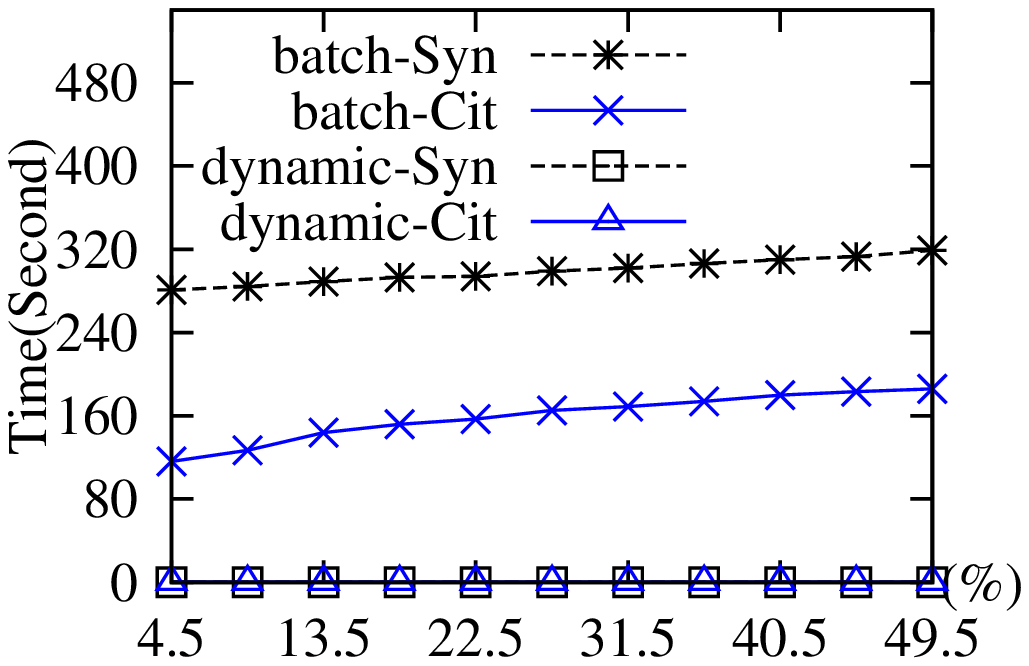}}
\hspace{0.2ex}
\subfigure[{\scriptsize Varying $|\Delta P|$ (capacity)}]{\label{fig-exp-patinc-cap}
\includegraphics[scale=0.38]{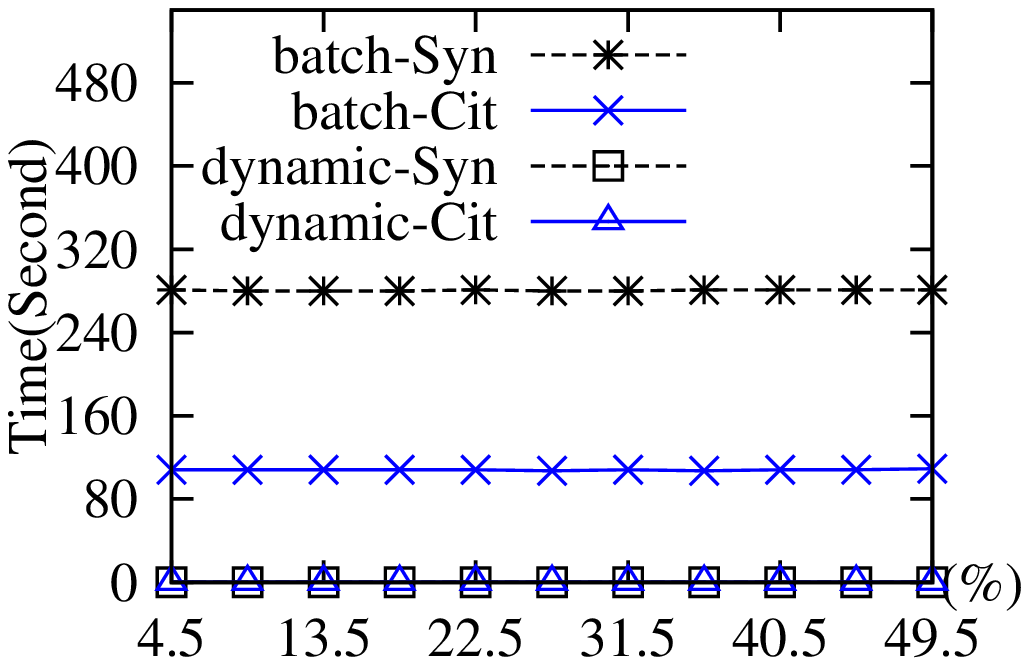}}
\hspace{0.2ex}
\subfigure[{\scriptsize Varying $|\Delta P|$ (hybrid updates)\hspace{-8.5ex}}]{\label{fig-exp-patinc-hyb}
\includegraphics[scale=0.38]{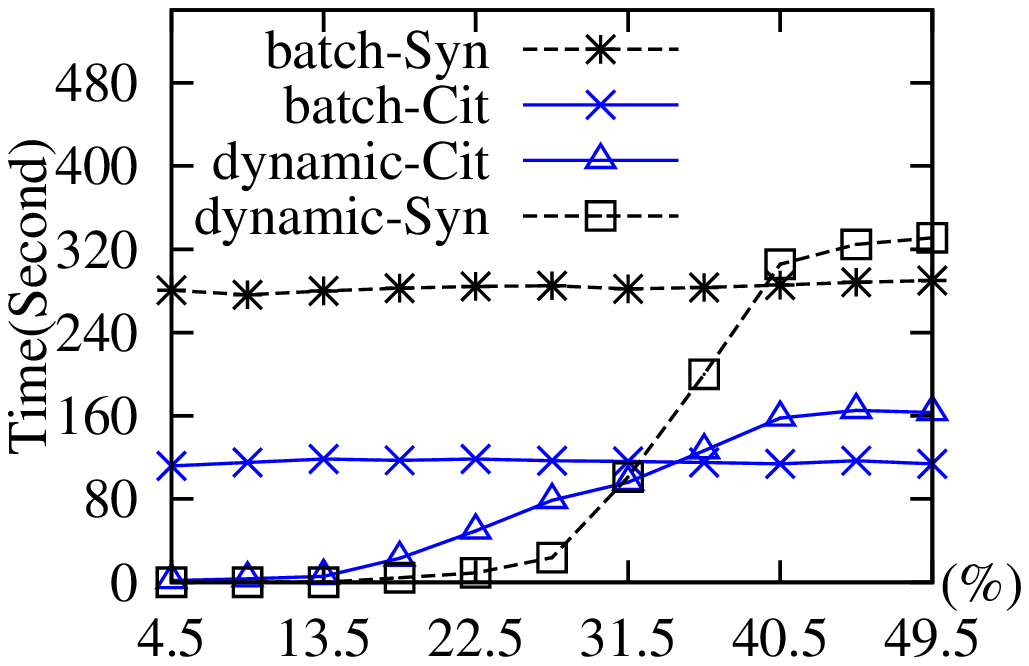}}
\vspace{-2.5ex}

\subfigure[{\scriptsize Varying $|\Delta G|$ (deletions)}]{\label{fig-exp-datainc-del}
\includegraphics[scale=0.38]{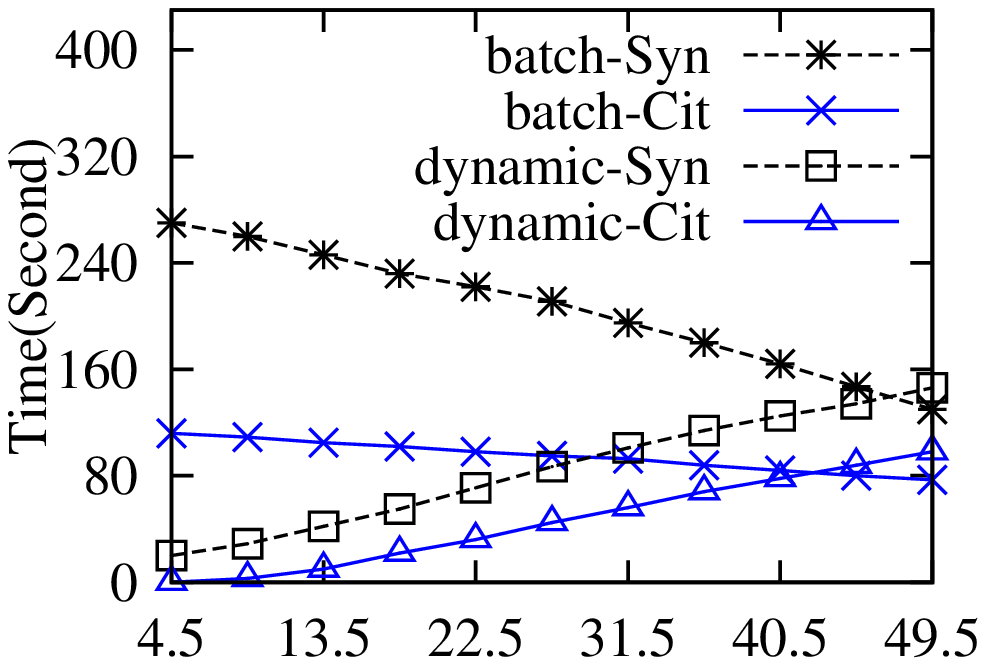}}
\hspace{0.2ex}
\subfigure[{\scriptsize Varying $|\Delta G|$ (insertions)}]{\label{fig-exp-datainc-ins}
\includegraphics[scale=0.38]{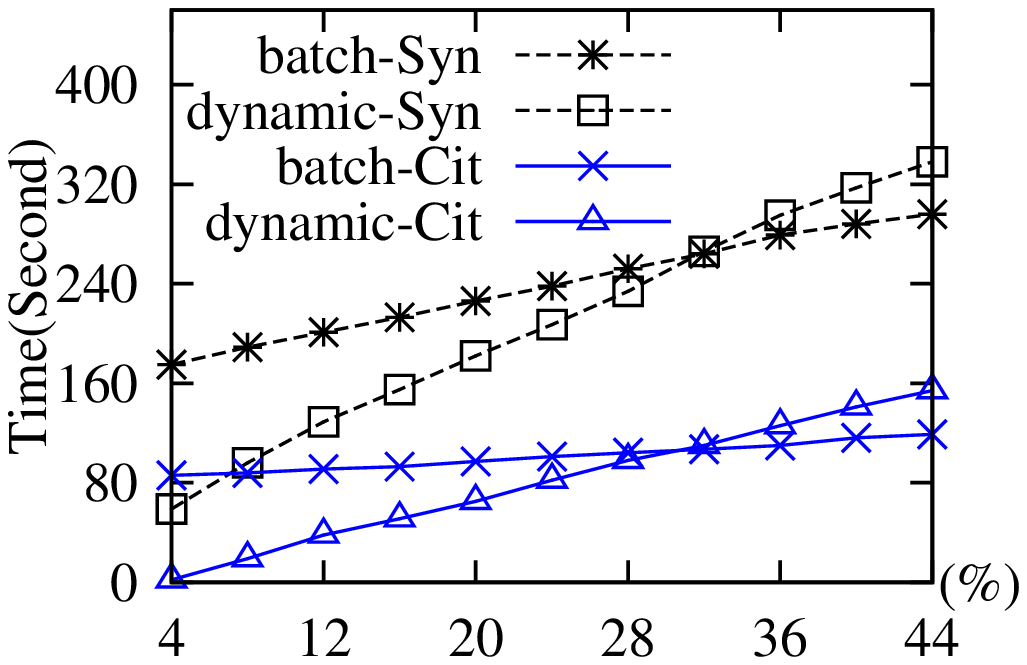}}
\hspace{0.2ex}
\subfigure[{\scriptsize Varying $|\Delta G|$ (hybrid updates)}\hspace{-5.5ex}]{\label{fig-exp-datainc-hyb}
\includegraphics[scale=0.38]{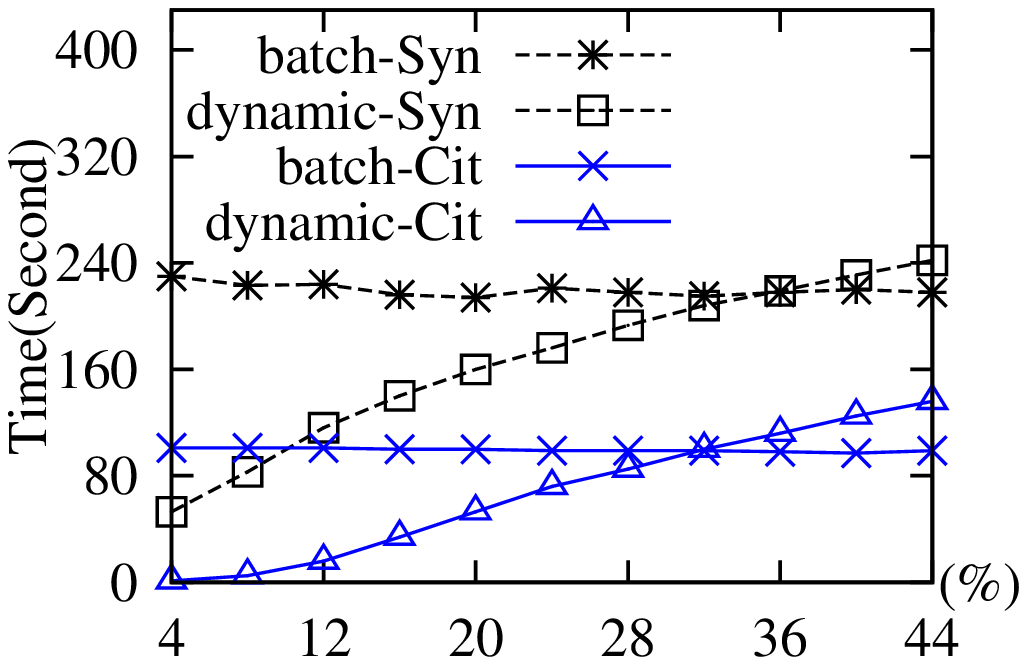}}
\hspace{0.2ex}
\subfigure[{\scriptsize Vary $(|\Delta P|,|\Delta G|)$ (simultaneous)}\hspace{-8.5ex}]{\label{fig-exp-hyb-patdata}
\includegraphics[scale=0.38]{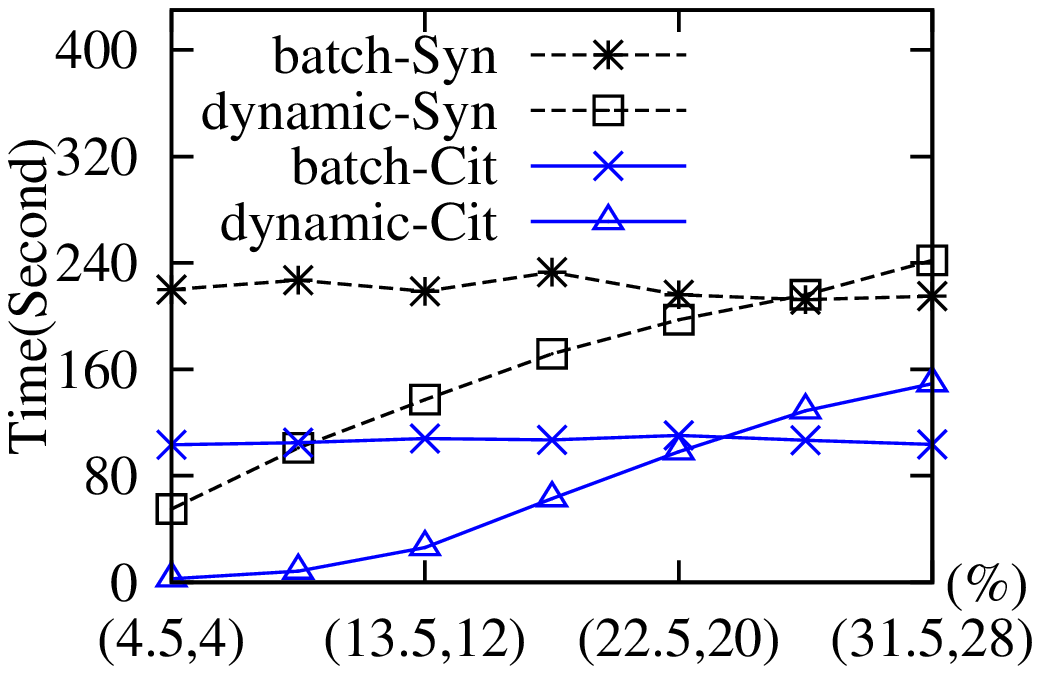}}
\vspace{-2.5ex}

\subfigure[{\scriptsize Varying $|\Delta P|$ in update sets}]{\label{fig-exp-patinc-multi}
\includegraphics[scale=0.38]{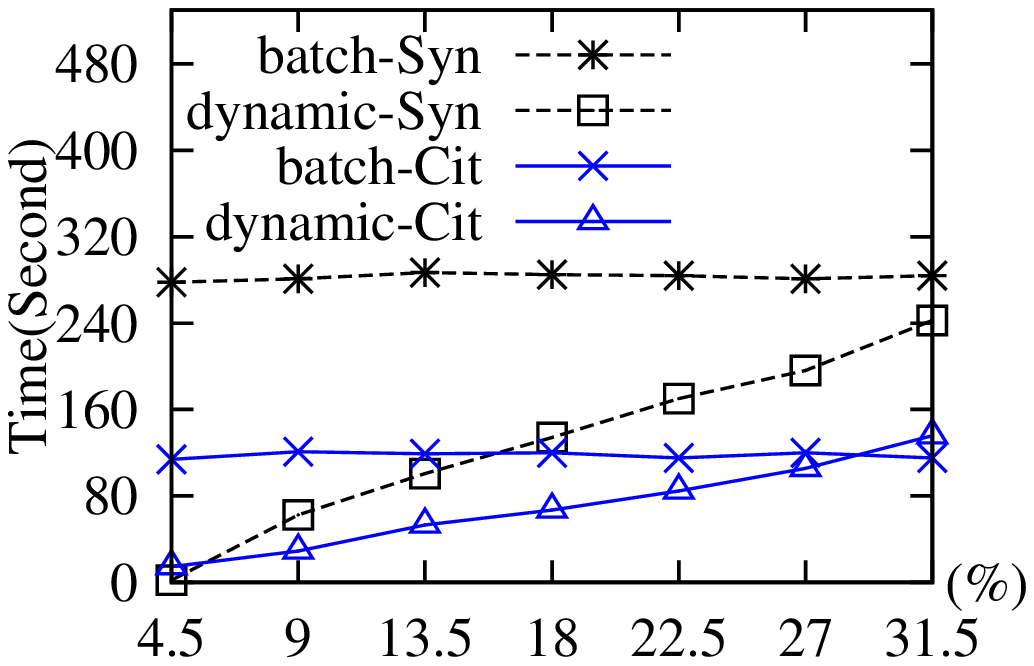}}
\hspace{0.2ex}
\subfigure[{\scriptsize Varying $|\Delta G|$ in update sets}]{\label{fig-exp-datainc-multi}
\includegraphics[scale=0.38]{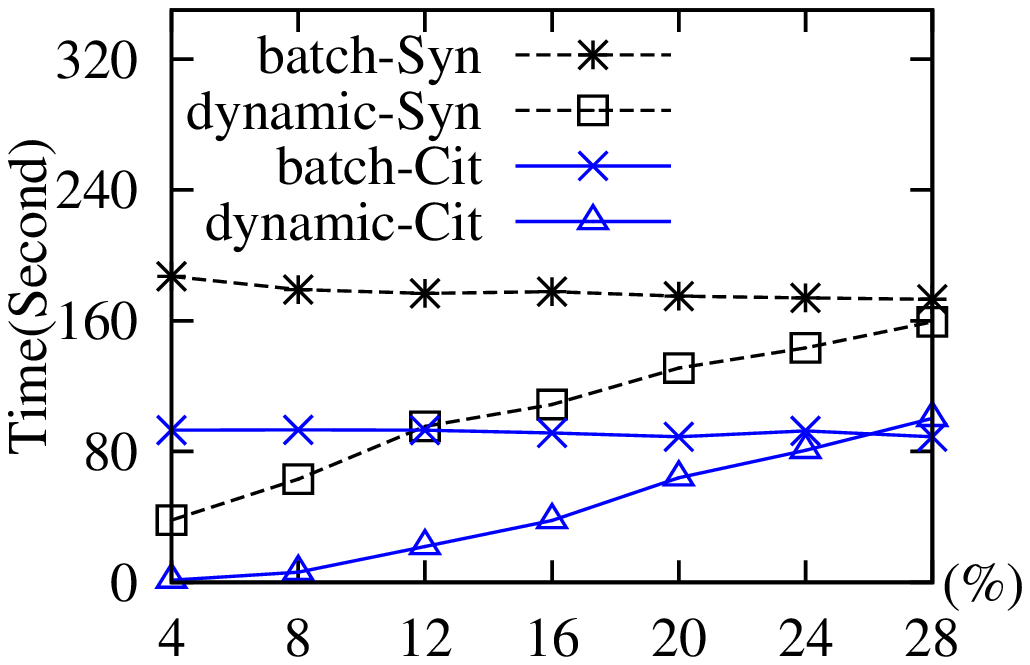}}
\hspace{0.2ex}
\subfigure[{\scriptsize Vary $(|\Delta P|,|\Delta G|)$ in update sets}\hspace{-8.5ex}]{\label{fig-exp-hyb-datapat-multi}
\includegraphics[scale=0.38]{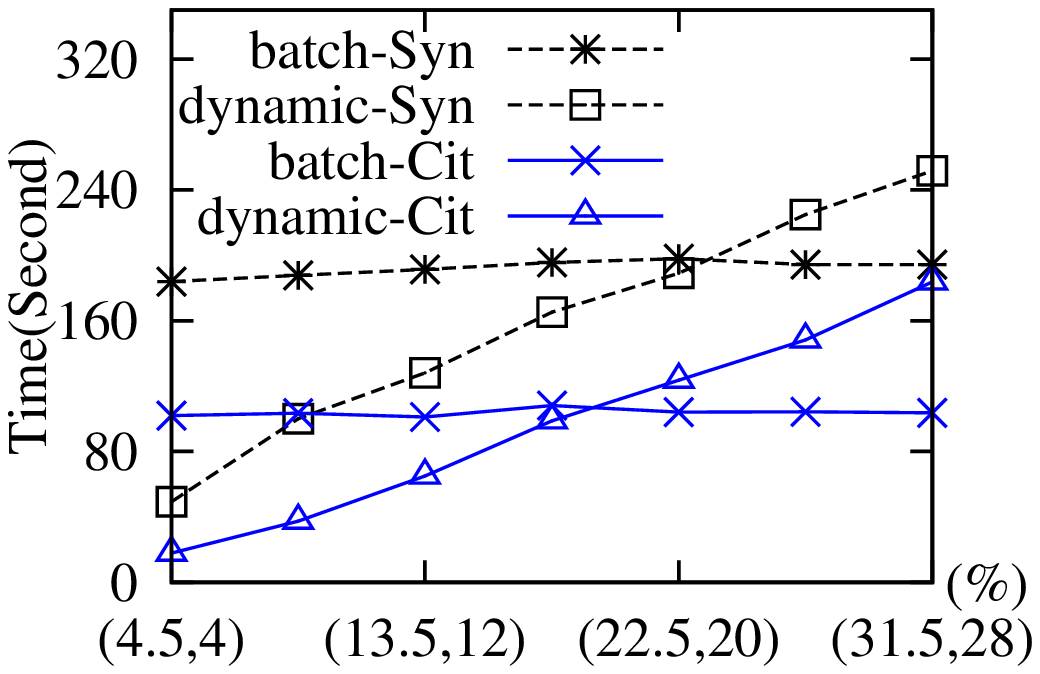}}
\hspace{0.2ex}
\subfigure[{\scriptsize Varying Datasets}]{\label{fig-exp-extraspace}
\includegraphics[scale=0.38]{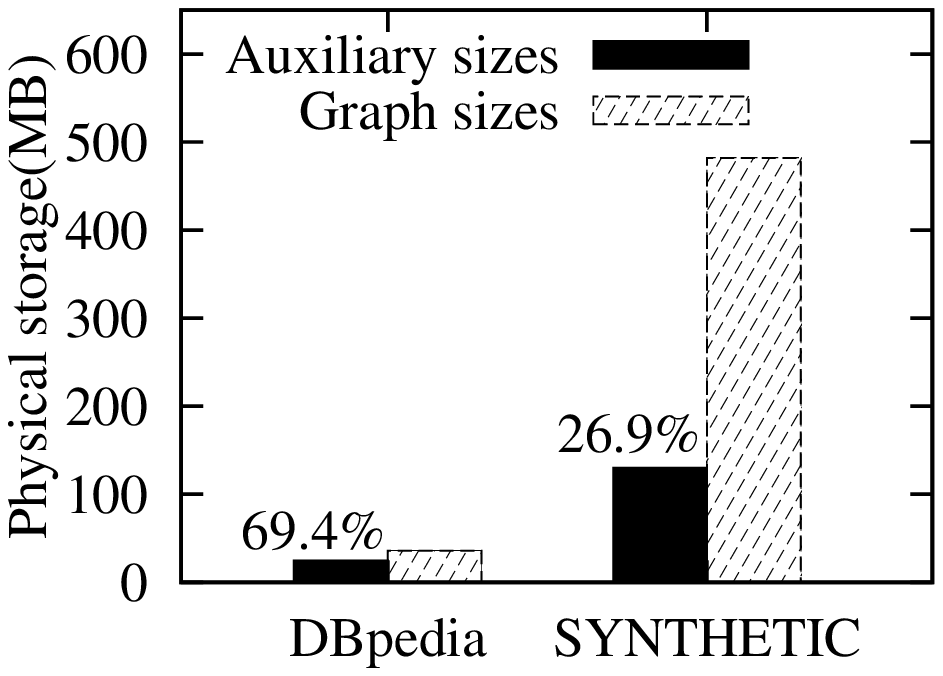}}
\vspace{-2ex}
\end{center}
\vspace{-2.5ex}
\caption{Performance evaluation of algorithm \inc for dynamic top-$k$ team formation (Cit: \citationd, Syn: \synthetic)}
\label{exp-inc}
\vspace{-2.5ex}
\end{figure*}

\stitle{Exp-3: Efficiency of \inc for single set of updates}. We evaluated the efficiency of algorithm \inc for processing one set of pattern updates, data updates and simultaneous pattern and data updates vs. algorithm \optgrouprec on \citationd and \synthetic, respectively.

\etitle{(i) Pattern updates}. We fixed $(|V_{P}|,$ $|E_{P}|)$ to be $(10, 12)$, and varied the number $|\Delta P|$ of unit updates from 1 to 11, corresponding to 4.5\% to 49.5\% in
Figs. \ref{fig-exp-patinc-del}, \ref{fig-exp-patinc-ins}, \ref{fig-exp-patinc-cap} and \ref{fig-exp-patinc-hyb},
which show the results when $\Delta P$ contains (edge and node) deletions, (edge and node) insertions, capacity changes and hybrid pattern updates (5 types) respectively,
while keeping the proportion for each type equal.

We find the following.
(1) \inc outperforms \optgrouprec even when deletions are no more than 40.5\% on \citationd and 49.5\% on \synthetic;
\inc consistently does better than \optgrouprec due to the early-return strategy.
(2) \inc improves \optgrouprec to a large extent when only processes insertions and capacity changes.
(3) For the same $|\Delta P|$, \inc needs less time to process insertions than deletions.
(4) When processes hybrid pattern updates, \inc outperforms \optgrouprec when changes are no more than (31.5\%, 40.5\%) on (\citationd, \synthetic);
It is because all balls are identified as \affballsx when pattern updates accumulate to a certain extent.

\etitle{(ii) Data updates}. For (edge and node) deletions (resp. insertions) on datasets,
\eg\ \citationd with $|G|=4.4M$,
we varied $|G|$ from $4.4M$ to $2.22M$ (resp. from $3.05M$ to $4.4M$) in $4.5\%$ decrements (resp.\,$4\%$ increments) by randomly picking a subset of nodes and edges and removing from $G$ (resp. inserting into $G$);
For hybrid data updates (4 types), we randomly sampled a subgraph $G_s$ and removed from $G$, obtaining the initial $G$. 
We varied $|G|$ by firstly removing a subset of nodes and edges from $G$ and then inserting a subset of nodes and edges from $G_s$ into $G$, in total  $4\%$ updates.
The results are shown in Figures \ref{fig-exp-datainc-del}, \ref{fig-exp-datainc-ins} and \ref{fig-exp-datainc-hyb}.

We find the following.
(1) \inc outperforms \optgrouprec when insertions are no more than 28\% and 32\% on \citationd and \synthetic (resp. 40.5\% and 45\% for deletions).
(2) For the same $|\Delta G|$, \inc needs less time to process deletions than insertions.
(3) We have conducted a survey: the user increment on Facebook~\cite{facebookStat} and Twitter~\cite{twitterStat} daily reaches 1.23\textperthousand\ and 2.47\textperthousand.
Therefore, \inc is able to handle the increments accumulated in dozens of days on Facebook and Twitter
at a high efficiency.
(4) \inc outperforms \optgrouprec when hybrid data updates are no more than 32\% and 36\% on \citationd and \synthetic, respectively.

\etitle{(iii) Simultaneous pattern and data updates}.  Varying the number of hybrid pattern updates from 1 to 7 and the amount of hybrid data updates from 4\% to 28\% together, corresponding to (4.5\%, 4\%) to (31.5\%, 28\%) for $(\Delta P, \Delta G)$ in Fig.~\ref{fig-exp-hyb-patdata}. We find that \inc outperforms \optgrouprec when $(\Delta P, \Delta G)$ is no more than (22.5\%, 20\%) and (27\%, 24\%) on \citationd and \synthetic, respectively.

\stitle{Exp-4: Efficiency of \inc for continuous sets of updates}. We evaluated \inc for a serial sets of pattern updates, data updates and simultaneous pattern and data updates vs. \optgrouprec on \citationd and \synthetic.

\etitle{(i) Pattern updates}. We generated $5$ sets of hybrid pattern updates, varying the number of updates in each set from $1$ to $7$.
We tested the average time took by \inc to finish all these sets one by one.
The results are reported in Fig.~\ref{fig-exp-patinc-multi}.

Recall that \inc adopts a lazy update policy, which definitely affects the processing time of next updates.
However, we find \inc outperforms \optgrouprec \wrt average time, when the amount of hybrid  pattern updates in each set is no more than (27\%, 31.5\%)  on (\citationd, \synthetic). This verifies the effectiveness of our lazy update policy.

\etitle{(ii) Data updates}. The setting is same as above. Varying the amount of hybrid data updates in each set from 4\% to 28\%,
the results are reported in Fig.~\ref{fig-exp-datainc-multi}.
We find that \inc outperforms \optgrouprec when hybrid data updates are no more than (24\%, 28\%) on (\citationd, \synthetic).

\etitle{(iii) Simultaneous pattern and data updates}. Using the same setting and varying the simultaneous pattern and data updates $(\Delta P, \Delta G)$ from (4.5\%, 4\%) to (31.5\%, 28\%) in Fig.~\ref{fig-exp-hyb-datapat-multi},
We find that \inc outperforms \optgrouprec when updates in each set are no more than (18\%, 16\%) and (22.5\%, 20\%)
on \citationd and \synthetic, respectively.

\eat{
\begin{table}[h!]
\begin{center}
\begin{small}
\scriptsize
\begin{tabular}{|c|c|c|c|}
\hline
{\bf Datasets}       &  {\bf Auxiliary sizes}  &  {\bf Graph sizes}  &  {\bf Ratio} \\
\hline\hline
{\bf \citationd}            &  25MB       &        36MB    &     69.4\%   \\ \hline
{\bf \synthetic}       &  130MB       &        482MB      &     26.9\%   \\ \hline
\end{tabular}
\vspace{-1.5ex}
\end{small}
\caption{Extra space of auxiliary data structures}
\label{tab-extraspace}
\vspace{-5ex}
\end{center}
\end{table}
}

\stitle{Exp-5: Physical storage of auxiliary structures}. As shown in Fig.~\ref{fig-exp-extraspace}, the incremental algorithm \inc takes (25MB, 130MB) extra space to store all its auxiliary structures on (\citationd, \synthetic), while they need (36MB, 482MB) space to store themselves.
That is, the auxiliary structures are light-weight, and only take (69.4\%, 26.9\%) extra space compared with the original datasets.


\stitle{Summary}. From these tests, we find the following.

\sstab(1) Our graph pattern matching approach is effective at capturing the practical requirements of top-$k$ team formation.

\sstab(2) Our batch algorithm for  top-$k$ team formation is efficient,
\eg it only took 116s when $|V|=1.39M$ and $|V_P|=10$.

\sstab (3) Our incremental algorithm for dynamic top-$k$ team formation is able to process continuous pattern and data updates, separately and simultaneously,
and it is more promising than its batch counterpart,
even (a) when changes are 36\% for pattern updates, 34\% for data updates, and (25\%, 22\%) for simultaneous pattern and data updates on average,
and (b) when 29\% for continuous pattern updates, 26\% for continuous data updates and (20\%, 18\%) for continuously simultaneous pattern and data updates on average.

\section{Conclusion}
\label{sec-conclusion}

We have introduced a graph pattern matching approach for (dynamic) top-$k$ team formation problem.
We have proposed team simulation,
based on which we have developed a \eat{efficient}batch algorithm for top-k team formation.
We have also developed a unified incremental algorithm to handle continuous pattern and data updates, separately and simultaneously.
We have experimentally verified the effectiveness and efficiency of the batch and incremental algorithms.
\looseness=-1

A couple of topics are targeted for future work.  First, an interesting topic is to develop distributed algorithms for top-$k$ team formation.
Second, the study of dynamic algorithms for query updates is in its infancy,
and hence, an important topic is to develop such algorithms for various problems.

%

\balance
\bibliographystyle{abbrv}
\begin{small}
\bibliography{paper}
\end{small}


\end{document}